\documentclass[prd,aps,nofootinbib,floatfix,11pt]{revtex4}
\usepackage{amsmath,graphicx,epsfig,amssymb,dsfont,mathtools,}
\usepackage[usenames]{color}
\usepackage{ulem} 
\usepackage{bigstrut}
\usepackage{slashed}
\usepackage{multirow}
\usepackage{subfigure}

\allowdisplaybreaks

\begin{document}\title{W-exchange contribution to the decays $\Xi_{cc}^{++}\to\Xi_{c}^{+(\prime)}\pi^{+}$ using \\ light-cone sum rules }

\author{Yu-Ji Shi$^{1}$~\footnote{Email: shiyuji92@126.com}, Zhen-Xing Zhao~$^{2}$~\footnote{Email: zhaozx19@imu.edu.cn}, Ye Xing~$^{3}$~\footnote{Email: xingye$\_$guang@cumt.edu.cn}  and Ulf-G. Mei{\ss}ner~$^{1,4,5}$~\footnote{Email: meissner@hiskp.uni-bonn.de}}
\affiliation{$^1$ Helmholtz-Institut f\"ur Strahlen- und Kernphysik and Bethe Center \\ for Theoretical Physics,
  Universit\"at Bonn, 53115 Bonn, Germany\\
  $^{2}$ School of Physical Science and Technology, Inner Mongolia University, Hohhot 010021, China\\
  $^{3}$ School of Physics, China University of Mining and Technology, Xuzhou 221000, China\\
  $^4$ Institute for Advanced Simulation, Institut f{\"u}r Kernphysik and J\"ulich Center for Hadron Physics,
  Forschungszentrum J{\"u}lich, D-52425 J{\"u}lich, Germany\\
$^5$ Tbilisi State University, 0186 Tbilisi, Georgia}

\begin{abstract}
  We calculate the W-exchange contribution to the $\Xi_{cc}^{++}\to\Xi_{c}^{+(\prime)}\pi^{+}$ decay  using  light-cone sum rules. The two-particle light-cone distribution amplitudes of the pion are used as non-perturbative input for the sum rules calculation, and the perturbative kernel is calculated at the leading order. We obtain the corresponding decay branching fractions by combining our W-exchange amplitudes with the factorizable amplitudes given by various theoretical methods from the literature. It is shown that with the factorizable amplitudes from  heavy quark effective theory, we obtain the branching fraction ratio  ${\cal B}(\Xi_{cc}^{++}\to\Xi_{c}^{+\prime}\pi^{+})/{\cal B}(\Xi_{cc}^{++}\to\Xi_{c}^{+}\pi^{+})= 1.42\pm 0.78$, which is consistent with the experimental value of $1.41\pm 0.17\pm 0.1$.
\end{abstract}
\maketitle

\section{Introduction}

The conventional quark model predicted the existence of doubly-heavy baryons consisting of two heavy quarks (bottom or charm quarks) \cite{Gell-Mann:1964ewy,Zweig:1964jf,DeRujula:1975qlm,Jaffe:1975us,Ponce:1978gk,Fleck:1988vm}. After pursuing the doubly charmed baryons for decades, in 2017 the LHCb collaboration announced the  observation of  the lowest-lying state $\Xi_{cc}^{++}$ with mass $3620.6\pm1.5(\text{stat})\pm0.4(\text{syst})\pm0.3(\Xi_{c}^{+})\text{MeV}/c^2$~\cite{LHCb:2017iph}. This new baryon was observed via the decay channel $\Xi_{cc}^{++}\to \Lambda_{c}^{+}K^{-}\pi^{+}\pi^{+}$ which is consistent with the prediction given by Ref.~\cite{Yu:2017zst}. One year later, in 2018, a two-body decay channel $\Xi_{cc}^{++}\to\Xi_c^{+}\pi^{+}$ was observed~\cite{LHCb:2018pcs}, which further confirms the existence of this doubly charmed baryon.  Recently, the  LHCb collaboration has observed a similar  decay channel, $\Xi_{cc}^{++}\to\Xi_c^{+{\prime}}\pi^{+}$, and measured the branching fraction ratio \cite{LHCb:2022rpd}:
\begin{align}
\frac{{\cal B}(\Xi_{cc}^{++}\to\Xi_c^{+\prime}\pi^{+})}{{\cal B}(\Xi_{cc}^{++}\to\Xi_c^{+}\pi^{+})}\equiv\frac{{\cal B}^{\prime}}{{\cal B}}=1.41\pm 0.17\pm 0.1,
\end{align}
which means that the branching fraction of the decay into $\Xi_c^{+\prime}$ is larger than that into $\Xi_c^{+}$.

\begin{figure}
\begin{center}
\includegraphics[width=0.9\columnwidth]{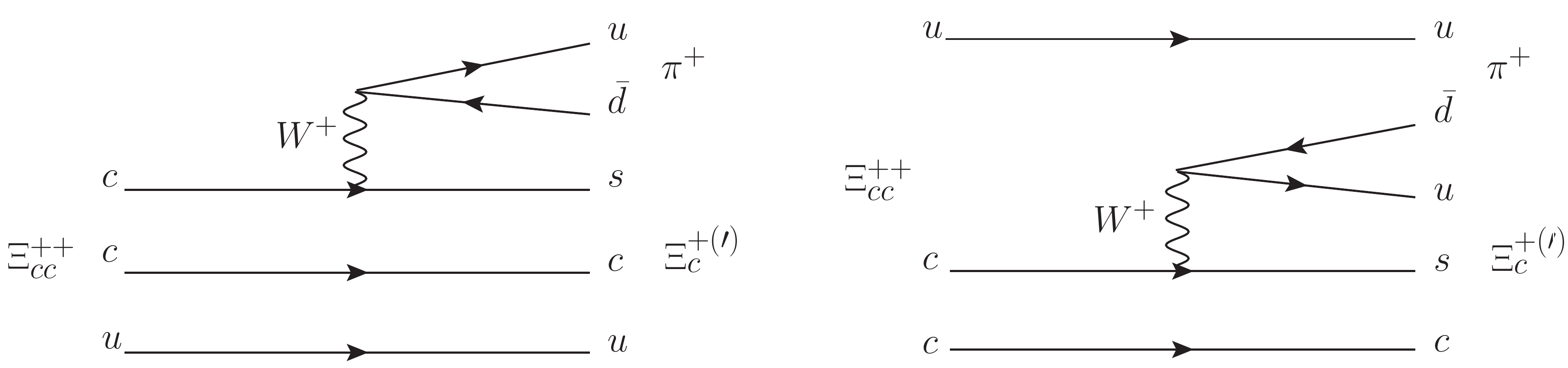} 
\caption{W-emission diagram (left) and W-exchange diagram (right) for the $\Xi_{cc}^{++}\to\Xi_{c}^{+(\prime)}\pi^{+}$ decay.}
\label{fig:XicctoXicpiTopo} 
\end{center}
\end{figure}
Theoretically, the weak decay $\Xi_{cc}^{++}\to\Xi_c^{+({\prime})}\pi^{+}$ receives  contributions from the two topological diagrams as shown in Fig.~\ref{fig:XicctoXicpiTopo}, the W-emission diagram (left) and the W-exchange diagram (right). Generally, according to the naive factorization \cite{Bjorken:1988kk,Wirbel:1985ji,Bauer:1986bm}, the W-emission diagram is approximately factorizable so that it can be calculated by considering the baryon transformation matrix element and the pion matrix element independently. For the decay of the $\Xi_{cc}^{++}$, the transformation matrix element is parameterized by form factors which have been evaluated in various theoretical works based on QCD sum rules (QCDSR) \cite{Shi:2019hbf}, light-cone sum rules (LCSR) \cite{Shi:2019fph,Hu:2019bqj,Hu:2022xzu,Aliev:2022tvs,Aliev:2022maw}, QCD factorization (QCDF) \cite{Sharma:2017txj,Gerasimov:2019jwp}, diquark effective theory (DiET) \cite{Shi:2020qde}, light-front quark model (LFQM) \cite{Wang:2017mqp,Zhao:2018mrg,Cheng:2020wmk,Ke:2019lcf,Ke:2022gxm,Hu:2020mxk}, constituent quark model (CQM) \cite{Gutsche:2018msz,Gutsche:2019iac}, non-relativistic quark model (NRQCD) and heavy quark effective theory (HQET) \cite{Sharma:2017txj,Dhir:2018twm}. Further, the pion matrix element can be  simply expressed by the pion decay constant. 

Unlike the weak decays of mesons, the W-exchange diagrams in the baryon decays are generally non-factorizable, and this difficulty increases  when we are facing the decays of doubly heavy baryons. Nowadays, except an SU(3) symmetry analysis \cite{Wang:2017azm,Shi:2017dto} and a phenomenological study \cite{Li:2017ndo},  there is no QCD based or 
model-independent study on the W-exchange contribution in doubly charmed baryon decays. The mostly used approach to evaluate such a contribution is the pole-model \cite{Sharma:2017txj,Gerasimov:2019jwp,Cheng:2020wmk,Gutsche:2018msz,Dhir:2018twm}. Recently, the combination of the  factorizable contribution and the non-factorizable contribution from the pole-model leads to the  ratio ${\cal B}^{\prime}/{\cal B}$ between 0.81 and 0.83 \cite{Sharma:2017txj,Gerasimov:2019jwp}, which means that the branching fraction of the decay into $\Xi_c^{+\prime}$ is smaller, which is obviously contrary to the experimental result. Furthermore, if one includes the interference between the W-emission and the W-exchange contributions, the ${\cal B}^{\prime}/{\cal B}$ will become much larger, namely $6.74$ \cite{Cheng:2020wmk}.

This deviation of the theoretical prediction from the experimental measurement implies that a more precise theoretical calculation for the W-exchange  contribution in the $\Xi_{cc}^{++}\to\Xi_c^{+({\prime})}\pi^{+}$ decay is necessary. In this work, we will use the method of LCSR to solve this problem. LCSR were firstly proposed to study the transition form factors of the radiative or semi-leptonic hadron decays \cite{Balitsky:1989ry,Braun:1988qv,Chernyak:1990ag}. In the framework of LCSR, the required transition matrix element can be extracted from a suitable correlation function at the hadron level. The quark-hadron duality enables us to relate this correlations function with  the one at the quark-gluon level,  where it can be calculated by the operator-product-expansion (OPE),  and all the non-perturbative contributions come from the light-cone distribution amplitudes (LCDAs) of a certain hadron in the decay.  After decades of development, a new technique of LCSR was proposed to study the  non-leptonic decay of the $B$ mesons into two light mesons \cite{Khodjamirian:2000mi,Khodjamirian:2003eq,Khodjamirian:2017zdu}. This new technique of LCSR can be extended to the case of heavy or doubly heavy baryon decays, and in this work we will use it to calculate the W-exchange  contribution in the $\Xi_{cc}^{++}\to\Xi_c^{+({\prime})}\pi^{+}$ decay.

This paper is organized as follows. In Sec.~\ref{sec:CorFunc_sum_rules}, we introduce a suitable correlation function to extract the decay amplitude of $\Xi_{cc}^{++}\to\Xi_c^{+({\prime})}\pi^{+}$. In Sec.~\ref{sec:Hdr_sum_rules}, we perform the hadron level calculation for the correlation function and extract the required decay amplitude. In Sec.~\ref{sec:QCD_sum_rules}, we perform the quark-gluon level calculation for the correlation function with the use of two-particle LCDAs of the pion. In Sec.~\ref{sec:numericalResult}, we give the numerical results on the decay amplitudes and branching fractions of $\Xi_{cc}^{++}\to\Xi_c^{+({\prime})}\pi^{+}$ and compare our results with those from the literature. Sec.~\ref{sec:conclusion} contains a brief summary of this work.

\section{The correlation function in LCSR}
\label{sec:CorFunc_sum_rules}

In this section, we give a suitable correlation function for the study of  the $\Xi_{cc}^{++}\to\Xi_{c}^{+(\prime)}\pi^{+}$ decay in the framework of the LCSR. The relevant effective Hamiltonian for this decay is
\begin{align}
{\cal H}_{\rm eff}&=\frac{G_{F}}{\sqrt{2}}V_{cs}V_{ud}^{*}\left(C_{1}{\cal O}_{1}+C_{2}{\cal O}_{2}\right),\nonumber\\
{\cal O}_{1}&=\bar{s}\gamma_{\mu}(1-\gamma_{5})c\ \bar{u}\gamma^{\mu}(1-\gamma_{5})d,\nonumber\\ 
{\cal O}_{2}&=\bar{s}_{a}\gamma_{\mu}(1-\gamma_{5})c_{b}\ \bar{u}_{b}\gamma^{\mu}(1-\gamma_{5})d_{a},\label{eq:effHamil}
\end{align}
where the $C_1, C_2$ are Wilson coefficients, and the subscripts $a, b$ are color indices. Generally, the transition matrix element of the  $\Xi_{cc}^{++}\to\Xi_{c}^{+(\prime)}\pi^{+}$ induced by ${\cal O}_{1,2}$ can be parameterized as
\begin{align}
\langle\Xi_{c}^{+(\prime)}(p-q)\pi^{+}(q)|{\cal O}_{i}(0)|\Xi_{cc}^{++}(p)\rangle=i\  \bar{u}(p-q)(A^{(\prime)i}+B^{(\prime)i}\gamma_{5})u(p).\label{eq:Heff}
\end{align}
Since the initial and final states are on-shell, $A^{(\prime)i}$ and $B^{(\prime)i}$ are just constants. In this work, our main task is to obtain the W-exchange contribution to  $A^{(\prime)i}$ and $B^{(\prime)i}$, denoted as $A^{(\prime)i}_{\rm WE}$ and $B^{(\prime)i}_{\rm WE}$ in what follows.

Using  the LCSR to calculate a transition matrix element, one begins with an appropriate correlation function which will be calculated both at the hadron and  the quark-gluon level. In our case, the correlation function corresponding to the $\Xi_{cc}^{++}\to\Xi_{c}^{+(\prime)}\pi^{+}$ decay is chosen as
\begin{align}
\Pi^{{\cal O}_{i}}(p,q,k)=i^{2}\int d^{4}x\ e^{-i(p-q)\cdot x}\int d^{4}y\ e^{i(p-k)\cdot y}\langle0|T\left\{ J_{\Xi_{c}^{(\prime)}}(y){\cal O}_{i}(0)\bar{J}_{\Xi_{cc}}(x)\right\} |\pi^{-}(q)\rangle,\label{eq:corrFunc}
\end{align}
where the hadron currents are defined as \cite{Shi:2019hbf}
\begin{align}
J_{\Xi_{c}} & =\frac{1}{\sqrt{2}}\varepsilon_{abc}\left(u_{a}^{T}C\gamma_{5}s_{b}-s_{a}^{T}C\gamma_{5}u_{b}\right)Q_{c},\nonumber\\
J_{\Xi_{c}^{\prime}} & =\frac{1}{\sqrt{2}}\varepsilon_{abc}\left(u_{a}^{T}C\gamma^{\mu}s_{b}+s_{a}^{T}C\gamma^{\mu}u_{b}\right)\gamma_{\mu}\gamma_{5}Q_{c},\nonumber\\
J_{\Xi_{cc}} & =\varepsilon_{abc}\left(Q_{a}^{T}C\gamma^{\mu}Q_{b}\right)\gamma_{\mu}\gamma_{5}u_{c}.
\end{align}
Following Ref.~\cite{Khodjamirian:2000mi}, we have temporarily included the pion in the initial state instead of the final state as in the real decay. The advantage of this  is to enable us to factorize out the matrix element $\langle 0|J_{\Xi_{c}^{\prime}}|\Xi_{c}^{+(\prime)}\rangle$ without any ambiguity after inserting the $\Xi_{c}^{+(\prime)}$ state into the right-hand side of $J_{\Xi_{c}^{\prime}}$. Otherwise, the final pion state has to be moved to the right over  $J_{\Xi_{c}}$ firstly before this factorization can be performed. Actually, during the calculation at the hadron level, by suitable analytic continuation this initial pion state will be moved to the final state.  We have also introduced an auxiliary momentum $k$ in the correlation function. This momentum is unphysical and should be set to zero at the end of the calculation. The reason why it must be introduced is closely related to the analytic continuation mentioned above, which will be explained in the next section.

\section{Hadron level calculation in  the LCSR}
\label{sec:Hdr_sum_rules}

In this section, we derive the calculation of the correlation function in Eq.~(\ref{eq:corrFunc}) at the hadron level. For simplicity, we take the case of the $\Xi_{c}^{+}$ as example, the derivation for the decays into $\Xi_{c}^{+\prime}$ is similar. 
We insert a complete set of states with the same quantum numbers as $J_{\Xi_{c}}(y)$ into the correlation function,
\begin{align}
\Pi_H^{{\cal O}_{i}}(p,q,k)_{\rm WE} =&\  i^{2}\int d^{4}xd^{4}y\ e^{-i(p-q)\cdot x}e^{i(p-k)\cdot y}\nonumber\\
 & \times\sum_{\pm^{\prime},\sigma^{\prime}}\int\frac{d^{3}\vec{l}}{(2\pi)^{3}}\frac{1}{2E_{l}}\langle0|J_{\Xi_{c}}(y)|l, \sigma^{\prime},\pm^{\prime}\rangle\langle l,\sigma^{\prime},\pm^{\prime}|{\cal O}_{i}(0)\bar{J}_{\Xi_{cc}}(x)|\pi^{-}(q)\rangle\nonumber\\
 & +\int_{s_{\Xi_{c}}}^{\infty}ds^{\prime}\frac{\rho_{\Xi_{c}}(s^{\prime},(p-q)^{2},P^{2})}{s^{\prime}-(p-k)^{2}},\label{eq:corrFuncInsertXic}
\end{align}
where the integration over $\rho_{\Xi_{c}}$ represents the contribution from the continuous  spectrum, $\pm^{\prime}$ corresponds to the
positive or negative parity of the $\Xi_{c}$ states, namely $\Xi_{c}(\frac{1}{2}^\pm)$, and $\sigma^{\prime}$ denotes the spin of the $\Xi_{c}$. The momentum of the one-particle state in the complete set $l$ should be on-shell, 
$l^{2}=m_{\Xi_{c}}^{\pm^{\prime}2}$. This means that the first matrix element in Eq.~(\ref{eq:corrFuncInsertXic}) can be simply parameterized by the $\Xi_{c}$ decay constants $\lambda_{\Xi_{c}}^{\pm}$:
\begin{align}
\langle0|J_{\Xi_{c}}(y)|l,\sigma^{\prime},+\rangle & =\lambda_{\Xi_{c}}^{+}u(l,\sigma^{\prime})e^{-il\cdot y},\nonumber\\
\langle0|J_{\Xi_{c}}(y)|l,\sigma^{\prime},-\rangle & =\lambda_{\Xi_{c}}^{-}i\gamma_{5}u(l,\sigma^{\prime})e^{-il\cdot y},\label{XicDecayConst}
\end{align}
while the second matrix element in Eq.~(\ref{eq:corrFuncInsertXic}) is a function of $p^{2},\ q^{2},\ k^{2},\ (p-q)^{2}$
and $P^{2}=(p-k-q)^{2}$. Generally speaking, a matrix element like this should also depend on $l^{2}$, however, the on-shell condition has reduced such a dependence. Further, since one can replace the $d^{3}\vec{l}$ integration by a four-momentum integration with the use of residue theorem,
\begin{align}
\int\frac{d^{3}\vec{l}}{(2\pi)^{3}}\frac{1}{2E_{l}}|l\rangle\langle l|\ e^{-il\cdot y}=\int\frac{d^{4}l}{(2\pi)^{4}}e^{-il\cdot y}\frac{1}{l^{2}-m_{\Xi_{c}}^{\pm^{\prime}2}}|l\rangle\langle l|,
\end{align}
after integrating over $d^{4}y$, one arrives at
\begin{align}
\Pi_H^{{\cal O}_{i}}(p,q,k)_{\rm WE} = &\ i^{3}\int d^{4}x\ e^{-i(p-q)\cdot x}\sum_{\pm^{\prime},\sigma^{\prime}}\frac{1}{(p-k)^{2}-m_{\Xi_{c}}^{\pm^{\prime}2}}\nonumber\\&\times\lambda_{\Xi_{c}}^{\pm^{\prime}}u^{\pm^{\prime}}(p-k,\sigma^{\prime})\langle p-k,\sigma^{\prime},\pm^{\prime}|{\cal O}_{i}(0)\bar{J}_{\Xi_{cc}}(x)|\pi^{-}(q)\rangle\nonumber\\
 & +\int_{s_{\Xi_{c}}}^{\infty}ds^{\prime}\frac{\rho_{\Xi_{c}}(s^{\prime},(p-q)^{2},P^{2})}{s^{\prime}-(p-k)^{2}},
\end{align}
where we have defined $u^{+}=u,\ u^{-}=i\gamma_{5}u$. To simplify the calculation we have chosen $p^{2}=k^{2}=0$ and $q^{2}=m_{\pi}^2\approx0$.
Now the correlation function depends on three Lorentz invariants $(p-k)^{2}$, $(p-q)^{2}$ and $P^{2}$, while the matrix element 
$\langle p-k,\sigma^{\prime},\pm^{\prime}|{\cal O}_{i}(0)\bar{J}_{\Xi_{cc}}(x)|\pi^{-}(q)\rangle$ only depends on $(p-q)^{2}$ and $P^{2}$. 

Further, the same correlation function can be calculated at the quark-hadron level by the OPE in the deep Euclidean region, $(p-k)^2\sim(p-q)^2\sim P^2\ll0$, which will be explicitly done in the next section. In principle, the expression of the same correlation function at these two levels should be equivalent:
\begin{align}
\Pi_H^{{\cal O}_{i}}(p,q,k)_{\rm WE}=\Pi_{QCD}^{{\cal O}_{i}}(p,q,k)_{\rm WE}=\frac{1}{\pi}\int_{(m_{c}+m_{s})^{2}}^{\infty}ds^{\prime}\frac{{\rm Im}\Pi_{QCD}^{{\cal O}_{i}}(s^{\prime},(p-q)^{2},P^{2})_{\rm WE}}{s^{\prime}-(p-k)^{2}},
\end{align}
where the $(p-k)^2$ dependence of $\Pi_{QCD}^{{\cal O}_{i}}$ has been described as in the form of a dispersion integral. We have omitted the $u$ quark mass and $(m_{c}+m_{s})^{2}$ is the quark level threshold to produce a $\Xi_c$ baryon. According to the quark-hadron duality, the integration over the continuous  spectrum at the hadron level is canceled by the corresponding integration above a certain threshold $s_{\Xi_{c}}$ at the quark-gluon level, which leads to the  equation:
 \begin{align}
 & i^{3}\int d^{4}x\ e^{-i(p-q)\cdot x}\sum_{\pm^{\prime},\sigma^{\prime}}\frac{1}{(p-k)^{2}-m_{\Xi_{c}}^{\pm^{\prime}2}}\lambda_{\Xi_{c}}^{\pm^{\prime}}u^{\pm^{\prime}}(p-k,\sigma^{\prime})\langle p-k,\sigma^{\prime},\pm^{\prime}|{\cal O}_{i}(0)\bar{J}_{\Xi_{cc}}(x)|\pi^{-}(q)\rangle_{\rm WE}\nonumber\\
= & \frac{1}{\pi}\int_{(m_{c}+m_{s})^{2}}^{s_{\Xi_{c}}}ds^{\prime}\frac{{\rm Im}\Pi_{QCD}^{{\cal O}_{i}}(s^{\prime},(p-q)^{2},P^{2})_{\rm WE}}{s^{\prime}-(p-k)^{2}}.\label{eq:corrFuncQHdual}
\end{align}
Generally, this threshold parameter should be slightly larger than the mass squared of the
corresponding hadron state. Here, it is chosen to be the same as in Ref.~\cite{Shi:2019hbf}, where
a QCDSR was used to study the semi-leptonic decay of the $\Xi_{cc}\to \Xi_{c}$.  It should be
mentioned that in principle $s_{\Xi_{c}}$ is a universal parameter which is process-independent
so that this procedure should be reasonable.

Note that since ${\rm Im}\Pi_{QCD}^{{\cal O}_{i}}(s^{\prime},(p-q)^{2},P^{2})_{\rm WE}$ is an analytic function
of $P^{2}$, the left-hand side of Eq.~(\ref{eq:corrFuncQHdual}) shares this property. 
Therefore one can extend $P^{2}$
to the physical region, namely $P^{2}>0$, which allows us to replace
the initial state $|\pi^{-}(q)\rangle$ of the matrix element on the
left-hand side by a final state $\langle\pi^{+}(-q)|$. After that,
a Borel transformation for $(p-k)^{2}$ on  both sides
of Eq.~(\ref{eq:corrFuncQHdual}) is performed, and we obtain
\begin{align}
 & -i^{3}\int d^{4}x\ e^{-i(p-q)\cdot x}\sum_{\pm^{\prime},\sigma^{\prime}}e^{-m_{\Xi_{c}}^{\pm^{\prime}2}/T^{\prime2}}\lambda_{\Xi_{c}}^{\pm^{\prime}}u^{\pm^{\prime}}(p-k,\sigma^{\prime})\langle p-k,\sigma^{\prime},\pm^{\prime};\pi^{+}(-q)|{\cal O}_{i}(0)\bar{J}_{\Xi_{cc}}(x)|0\rangle_{\rm WE}\nonumber\\
= & \frac{1}{\pi}\int_{(m_{c}+m_{s})^{2}}^{s_{\Xi_{c}}}ds^{\prime}e^{-s^{\prime}/T^{\prime2}}{\rm Im}\Pi_{QCD}^{{\cal O}_{i}}(s^{\prime},(p-q)^{2},P^{2})_{\rm WE}.\label{eq:corrFuncQHdualBorel1}
\end{align}
Next, we insert another complete set of states with the same quantum numbers as ${\bar J}_{\Xi_{cc}}$ 
into the left-hand side of Eq.~(\ref{eq:corrFuncQHdualBorel1}), which becomes
\begin{align}
-&\sum_{\pm^{\prime},\pm,\sigma^{\prime},\sigma}e^{-m_{\Xi_{c}}^{\pm^{\prime}2}/T^{\prime2}}\frac{1}{(p-q)^{2}-m_{\Xi_{cc}}^{\pm2}}\lambda_{\Xi_{c}}^{\pm^{\prime}}\lambda_{\Xi_{cc}}^{\pm}\nonumber\\
  &\times \  u^{\pm^{\prime}}(p-k,\sigma^{\prime})\langle p-k,\sigma^{\prime},\pm^{\prime};\pi^{+}(-q)|{\cal O}_{i}(0)|p-q,\sigma,\pm\rangle_{\rm WE}\ \bar{u}^{\pm}(p-q.\sigma)\nonumber\\
  &+\int_{s_{\Xi_{cc}}}^{\infty}ds\frac{\rho_{\Xi_{cc}}(s,P^{2})}{s-(p-q)^{2}}.
\end{align}
For the same reason as discussed before, the matrix element
$\langle p-k,s^{\prime},\pm^{\prime};\pi^{+}(-q)|{\cal O}_{i}(0)|p-q,s,\pm\rangle_{\rm WE}$
only depends on $P^{2}$. The $(p-q)^2$ dependence on the right-hand side of Eq.~(\ref{eq:corrFuncQHdualBorel1}) can
be further expressed as a dispersion integral,
\begin{align}
\frac{1}{\pi^{2}}\int_{(m_{c}+m_{s})^{2}}^{s_{\Xi_{c}}}ds^{\prime}e^{-s^{\prime}/T^{\prime2}}\int_{4m_{c}^{2}}^{\infty}ds\ \frac{1}{s-(p-q)^{2}}{\rm Im}^{2}\Pi_{QCD}^{{\cal O}_{i}}(s^{\prime},s,P^{2})_{\rm WE}~.
\end{align}
Now using the quark-hadron duality again to cancel out the $ds^{\prime}$ integration above a
certain threshold $s_{\Xi_{cc}}$ corresponding to the lowest $\Xi_{cc}$ state, and performing
the Borel transformation for $(p-q)^{2}$, we arrive at
\begin{align}
 & \sum_{\pm^{\prime},\pm,\sigma^{\prime},\sigma}e^{-m_{\Xi_{c}}^{\pm^{\prime}2}/T^{\prime2}-m_{\Xi_{cc}}^{\pm2}/T^{2}}\lambda_{\Xi_{c}}^{\pm^{\prime}}\lambda_{\Xi_{cc}}^{\pm}\nonumber\\
 &\times u^{\pm^{\prime}}(p-k,\sigma^{\prime})\langle p-k,\sigma^{\prime},\pm^{\prime};\pi^{+}(-q)|{\cal O}_{i}(0)|p-q,\sigma,\pm\rangle_{\rm WE}\ \bar{u}^{\pm}(p-q,\sigma)\nonumber\\
= & \frac{1}{\pi^{2}}\int_{(m_{c}+m_{s})^{2}}^{s_{\Xi_{c}}}ds^{\prime}\int_{4m_{c}^{2}}^{s_{\Xi_{cc}}}ds\ e^{-s^{\prime}/T^{\prime2}}e^{-s/T^{2}}{\rm Im}^{2}\Pi_{QCD}^{{\cal O}_{i}}(s^{\prime},s,P^{2}).\label{eq:sumruleEq}
\end{align}
Due to the existence of the auxiliary momentum $k$, unlike Eq.~(\ref{eq:Heff}),
the matrix element appearing in Eq.~(\ref{eq:sumruleEq}) must be parameterized by four terms:
\begin{align}
 & \langle p-k,\sigma^{\prime},\pm^{\prime};\pi^{+}(-q)|{\cal O}_{i}(0)|p-q,\sigma,\pm\rangle_{\rm WE}\nonumber\\
= &\  i\ \bar{u}^{\pm^{\prime}}(p-k,\sigma^{\prime})\left[A_{1,i}^{\pm^{\prime}\pm}(P^{2})+B_{1,i}^{\pm^{\prime}\pm}(P^{2})\gamma_{5}+A_{2,i}^{\pm^{\prime}\pm}(P^{2})\frac{\slashed q}{m_{\Xi_{cc}}^{\pm}}+B_{2,i}^{\pm^{\prime}\pm}(P^{2})\frac{\slashed q\gamma_{5}}{m_{\Xi_{cc}}^{\pm}}\right]u^{\pm}(p-q,\sigma).\label{eq:parMatrix1}
\end{align}
The form factors $A_{1,2,i}^{\pm^{\prime}\pm}\ \text{and}\ B_{1,2,i}^{\pm^{\prime}\pm}$  are functions of $P^{2}$.
Using the sum rules equation given in Eq.~(\ref{eq:sumruleEq}), we can extract these four functions.
Summing up the spin indices, we have
\begin{align}
 &i\  e^{-m_{\Xi_{c}}^{+2}/T^{\prime2}-m_{\Xi_{cc}}^{+2}/T^{2}}\lambda_{\Xi_{c}}^{+}\lambda_{\Xi_{cc}}^{+}(\slashed p_{2}+m_{\Xi_{c}}^{+})\left[A_{1,i}^{++}+B_{1,i}^{++}\gamma_{5}+A_{2,i}^{++}\frac{\slashed q}{m_{\Xi_{cc}}^{\pm}}+B_{2,i}^{++}\frac{\slashed q\gamma_{5}}{m_{\Xi_{cc}}^{\pm}}\right](\slashed p_{1}+m_{\Xi_{cc}}^{+})\nonumber\\
+ &i\  e^{-m_{\Xi_{c}}^{-2}/T^{\prime2}-m_{\Xi_{cc}}^{+2}/T^{2}}\lambda_{\Xi_{c}}^{-}\lambda_{\Xi_{cc}}^{+}(\slashed p_{2}-m_{\Xi_{c}}^{+})\left[A_{1,i}^{-+}+B_{1,i}^{-+}\gamma_{5}+A_{2,i}^{-+}\frac{\slashed q}{m_{\Xi_{cc}}^{\pm}}+B_{2,i}^{-+}\frac{\slashed q\gamma_{5}}{m_{\Xi_{cc}}^{\pm}}\right](\slashed p_{1}+m_{\Xi_{cc}}^{+})\nonumber\\
+ &i\  e^{-m_{\Xi_{c}}^{+2}/T^{\prime2}-m_{\Xi_{cc}}^{-2}/T^{2}}\lambda_{\Xi_{c}}^{+}\lambda_{\Xi_{cc}}^{-}(\slashed p_{2}+m_{\Xi_{c}}^{+})\left[A_{1,i}^{+-}+B_{1,i}^{+-}\gamma_{5}+A_{2,i}^{+-}\frac{\slashed q}{m_{\Xi_{cc}}^{\pm}}+B_{2,i}^{+-}\frac{\slashed q\gamma_{5}}{m_{\Xi_{cc}}^{\pm}}\right](\slashed p_{1}-m_{\Xi_{cc}}^{-})\nonumber\\
+ &i\  e^{-m_{\Xi_{c}}^{-2}/T^{\prime2}-m_{\Xi_{cc}}^{-2}/T^{2}}\lambda_{\Xi_{c}}^{-}\lambda_{\Xi_{cc}}^{-}(\slashed p_{2}-m_{\Xi_{c}}^{+})\left[A_{1,i}^{--}+B_{1,i}^{--}\gamma_{5}+A_{2,i}^{--}\frac{\slashed q}{m_{\Xi_{cc}}^{\pm}}+B_{2,i}^{--}\frac{\slashed q\gamma_{5}}{m_{\Xi_{cc}}^{\pm}}\right](\slashed p_{1}-m_{\Xi_{cc}}^{-})\nonumber\\
= & \frac{1}{\pi^{2}}\int_{(m_{c}+m_{s})^{2}}^{s_{\Xi_{c}}}ds^{\prime}\int_{4m_{c}^{2}}^{s_{\Xi_{cc}}}ds\ e^{-s^{\prime}/T^{\prime2}}e^{-s/T^{2}}{\rm Im}^{2}\Pi_{QCD}^{{\cal O}_{i}}(s^{\prime},s,P^{2})_{\rm WE},
\end{align}
with $p_{1}=p-q$ and $p_{2}=p-k$. Note that there are 16 independent spinor structures, and this
number is equal to that
of the form factors $A_{1,2}^{\pm^{\prime}\pm}\ \text{and}\ B_{1,2}^{\pm^{\prime}\pm}$. This matching enables
us to solve all of these form factors from the equation above. Finally, what we really care about
is the transition matrix element of positive parity baryons with $k=0$ and $P^{2}=m_{\Xi cc}^{+2}$.
Therefore, considering the $k\to0$ limit, and using the equation of motion, the matrix element
in Eq.~(\ref{eq:parMatrix1}) can be simplified as 
\begin{align}
 & \langle p-k,\sigma^{\prime},+;\pi^{+}(-q)|{\cal O}_{i}(0)|p-q,\sigma,+\rangle\mid_{k\to0,P^{2}=m_{\Xi cc}^{+2}}\nonumber\\
 =&\ i\  \bar{u}^{+}(p-k,\sigma^{\prime})\left\{ \left[A_{1,i}^{++}(m_{\Xi cc}^{+2})+\left(1-\frac{m_{\Xi_{c}}^{+}}{m_{\Xi_{cc}}^{+}}\right)A_{2,i}^{++}(m_{\Xi cc}^{+2})\right]\right.\nonumber\\
 &\left.+\left[B_{1,i}^{++}(m_{\Xi cc}^{+2})-\left(1+\frac{m_{\Xi_{c}}^{+}}{m_{\Xi_{cc}}^{+}}\right)B_{2,i}^{++}(m_{\Xi cc}^{+2})\right]\gamma_{5}\right\} u^{+}(p-q,\sigma)\nonumber\\
 \equiv &\ i\   \bar{u}^{+}(p,\sigma^{\prime})\left( A_{\rm WE}^i+B_{\rm WE}^i\gamma_{5}\right)u^{+}(p-q,\sigma).\label{eq:parMatrix0WE}
\end{align}
Now the four unknown form factors are reduced to two constants $A_{\rm WE}^i$ and $B_{\rm WE}^i$
($A_{\rm WE}^{\prime i}$ and $B_{\rm WE}^{\prime i}$ for the case of $\Xi_c^{\prime}$), which has
the same form as Eq.~(\ref{eq:parMatrix0}). 

\section{Quark-Gluon Level}
\label{sec:QCD_sum_rules}
\begin{figure}
\begin{center}
\includegraphics[width=0.5\columnwidth]{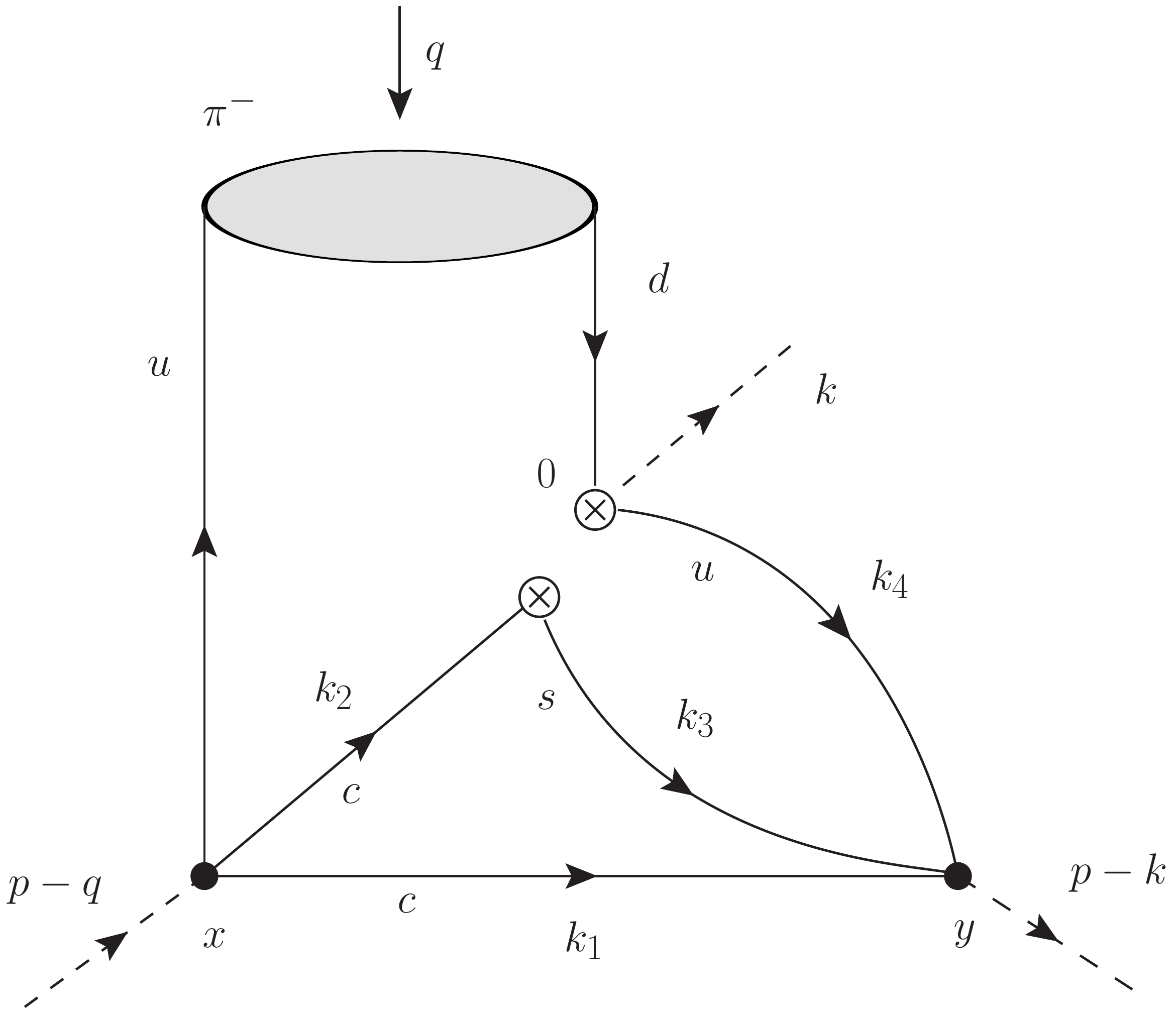} 
\caption{Feynman diagram of the W-exchange contribution to the correlation function in
  Eq.~(\ref{eq:corrFunc}). The gray bubble denotes the pion LCDAs, the black dots
  denote the baryon currents and the white double crossed dot represents the four-fermion
  interaction vertex from ${\cal O}_{1,2}$. The dashed lines with arrows denote the external
  momentum flows.}
\label{fig:CorreFuncWE} 
\end{center}
\end{figure}
In this section, we use the OPE to calculate the correlation function in Eq.~(\ref{eq:corrFunc})
at the quark-gluon
level. In the deep Euclidean region, $(p-k)^2\sim(p-q)^2\sim P^2\ll0$, the W-exchange contribution
to the correlation function can be expressed as a convolution of the perturbative kernel and
a nonperturbative matrix element of the pion:
\begin{align}
\Pi_{QCD}^{{\cal O}_{1}}(p,q,k)_{\alpha\sigma} & =-\Pi_{QCD}^{{\cal O}_{2}}(p,q,k)_{\alpha\sigma}\nonumber\\
 & =-2\sqrt{2}\varepsilon_{abc}\varepsilon_{ebc}\int d^{4}xd^{4}y\ e^{-i(p-q)\cdot x}e^{i(p-k)\cdot y}\nonumber\\
 & \times\left[S_{Q}(y-x)\gamma^{\nu}CS_{Q}^{T}(-x)C(1-\gamma_{5})\gamma_{\mu}CS_{s}^{T}(y)C\gamma_{5}S_{u}(y)\gamma^{\mu}(1-\gamma_{5})\right]_{\alpha\beta}(\gamma_{\nu}\gamma_{5})_{\rho\sigma}\nonumber\\
 & \times\langle0|\bar{u}_{e}^{\rho}(x)d_{a}^{\beta}(0)|\pi^{-}(q)\rangle,
\end{align}
where $\alpha,\beta,\rho,\sigma$ are spinor indices and $S_{Q,u,s}$  are the free propagators of
the $c, u, s$ quarks. The superscript "$T$" denotes the transposed in  spinor space.
The last matrix element can be expressed by the Light-Cone Distribution Amplitudes (LCDAs) of the
pion. Fig.~\ref{fig:CorreFuncWE} shows the Feynman diagram of the W-exchange effect in
the correlation function, where the gray bubble denotes the pion LCDAs.
The contribution from two-particle LCDAs of pion up to the twist-3 order are defined as
\cite{Duplancic:2008ix,Khodjamirian:2020mlb}
\begin{align}
\langle0|\bar{u}_{e}^{\rho}(x)d_{a}^{\beta}(0)|\pi^{-}(q)\rangle & =-\frac{i}{12}\delta_{ae}f_{\pi}\int_{0}^{1}du\ e^{-i\bar{u}q\cdot x}\Big[(\slashed p\gamma_{5})_{\beta\rho}\varphi_{\pi}(u)+(\gamma_{5})_{\beta\rho}\mu_{\pi}\phi_{3\pi}^{p}(u)\nonumber\\
 & +\frac{1}{6}(\gamma_{5}\sigma_{\mu\nu})_{\beta\rho}q^{\mu}x^{\nu}\mu_{\pi}\phi_{3\pi}^{\sigma}(u)\Big],
\end{align}
where
\begin{align}
\varphi_{\pi}(u)&=6 u \bar{u}\left(1+a_{2} C_{2}^{3 / 2}(u-\bar{u})+a_{4} C_{4}^{3 / 2}(u-\bar{u})\right),\nonumber\\
\phi_{3 \pi}^{p}(u)&=1+30 \frac{f_{3 \pi}}{\mu_{\pi} f_{\pi}} C_{2}^{1 / 2}(u-\bar{u})-3 \frac{f_{3 \pi} \omega_{3 \pi}}{\mu_{\pi} f_{\pi}} C_{4}^{1 / 2}(u-\bar{u}),\nonumber\\
\phi_{3 \pi}^{\sigma}(u)&=6 u(1-u)\left(1+5 \frac{f_{3 \pi}}{\mu_{\pi} f_{\pi}}\left(1-\frac{\omega_{3 \pi}}{10}\right) C_{2}^{3 / 2}(u-\bar{u})\right),
\end{align}
are the twist-2, twist-3p and twist-$3\sigma$ LCDAs respectively. $a_2=0.27$, $a_4=0.179$, $\mu_{\pi}=2.87$~GeV, $f_{3\pi}=0.0045$~GeV$^2$, $\omega_{\pi}=-1.5$ and
$f_{\pi}=0.13$~GeV  \cite{Duplancic:2008ix}. $C^{\alpha}_{n}$ are the Gegenbauer polynomials. Now
the derivation becomes straightforward, and we take the twist-2 LCDA as an example. Its
contribution to the correlation function is
\begin{align}
&\Pi_{QCD}^{{\cal O}_{1}}(p,q,k)_{(2)} \nonumber\\
 =& -4\sqrt{2}N_{c}\left(-\frac{i}{12}\right)f_{\pi}\int_{0}^{1}du\ \varphi_{\pi}(u)\int d^{4}xd^{4}y\int\frac{d^{4}k_{1}}{(2\pi)^{4}}\frac{d^{4}k_{2}}{(2\pi)^{4}}\frac{d^{4}k_{3}}{(2\pi)^{4}}\frac{d^{4}k_{4}}{(2\pi)^{4}}\nonumber\\
& \times e^{-i\bar{u}q\cdot x}e^{-ik_{1}\cdot(y-x)}e^{ik_{2}\cdot x}e^{-ik_{3}\cdot y}e^{-ik_{4}\cdot y}e^{-i(p-q)\cdot x}e^{i(p-k)\cdot y}\frac{1}{(k_{1}^{2}-m_{c}^{2})(k_{2}^{2}-m_{c}^{2})(k_{3}^{2}-m_{s}^{2})k_{4}^{2}}\nonumber\\
 & \times\left[(\slashed k_{1}+m_{c})\gamma^{\nu}(\slashed k_{2}-m_{c})(1-\gamma_{5})\gamma_{\mu}(\slashed k_{3}-m_{s})\gamma_{5}\slashed k_{4}\gamma^{\mu}(1-\gamma_{5})\slashed q\gamma_{5}\gamma_{\nu}\gamma_{5}\right]~, \label{eq:corrFuncMoment}
\end{align}
with $N_c$ the number of colors.

The double imaginary part of the correlation function is related to its double discontinuity,
which can be extracted by the cutting rules.
Setting the momentum of each propagator on-shell, we have
\begin{align}
&{\rm Im}^{2}\Pi_{QCD}^{{\cal O}_{1}}(s^{\prime},s,P^{2})_{(2)}=\frac{1}{(2i)^{2}}{\rm Disc^{2}}\Pi_{QCD}^{{\cal O}_{1}}(s^{\prime},s,P^{2})_{(2)} \nonumber\\
= & -4\sqrt{2}N_{c}\left(-\frac{i}{12}\right)f_{\pi}(-2\pi i)^{4}\frac{1}{(2i)^{2}}\frac{1}{(2\pi)}\int_{0}^{1}du\ \varphi_{\pi}(u)\int dm_{34}^{2}\int d\Phi_{\Delta}(P_{1}^{2},p_{2}^{2})\int d\Phi_{2}(m_{34}^{2})\nonumber\\
 & \times\left[(\slashed k_{1}+m_{c})\gamma^{\nu}(\slashed k_{2}-m_{c})(1-\gamma_{5})\gamma_{\mu}(\slashed k_{3}-m_{s})\gamma_{5}\slashed k_{4}\gamma^{\mu}(1-\gamma_{5})\slashed q\gamma_{5}\gamma_{\nu}\gamma_{5}\right].
\end{align}
Here we have introduced an extra integration on $m_{34}^2=(k_3+k_4)^2$ to express the 
double discontinuity of the correlation function as a convolution of a two-body phase space integration $\int d\Phi_{2}(m_{34}^{2})$ and a triangle diagram integration $\int d\Phi_{\Delta}(P_{1}^{2},p_{2}^{2})$,
\begin{align}
\int d\Phi_{2}(m_{34}^{2}) =&\ \int\frac{d^{3}k_{3}}{(2\pi)^{3}}\frac{1}{2E_{k_{3}}}\frac{d^{3}k_{4}}{(2\pi)^{3}}\frac{1}{2E_{k_{4}}}\delta^{4}(k_{34}-k_{3}-k_{4}),\ \ \ \ k_{34}^{2}=m_{34}^{2},\nonumber\\
\int d\Phi_{\Delta}(P_{1}^{2},p_{2}^{2}) =&\ \int d^{4}k_{34}d^{4}k_{1}d^{4}k_{2}\delta(k_{1}^{2}-m_{c}^{2})\delta(k_{2}^{2}-m_{c}^{2})\delta(k_{34}^{2}-m_{34}^{2})\nonumber\\
& \times\delta^{4}(P_{1}-k_{1}-k_{2})\delta^{4}(p_{2}-k_{1}-k_{34}),
\end{align}
where $P_{1}=p-uq=p_{1}+\bar{u}q$, and $p_1^2=s,\ p_2^2=s^{\prime 2}$. This factorization for the momentum
flows can also be understood intuitively by Fig.~\ref{fig:CorreFuncWE}. It is seen that the momentum
flowing into the left-lower corner of the triangle diagram is not $p_{1}$ itself
but $P_{1}=p-uq=p_{1}+\bar{u}q$ instead, and these momenta are related as $P_{1}^{2}=up_{1}^{2}=us$.  

For the contribution from the  twist-$3p$ and twist-$3\sigma$
LCDAs the calculation is similar. The only difference is that the twist-$3\sigma$ LCDA contains
a term proportional to the coordinate $x^{\nu}$. Note that from Eq.~(\ref{eq:corrFuncMoment})
there is an exponential term ${\rm exp}(iu q\cdot x)$ in the correlation function. One can
use it to express $x^{\nu}$ as $(-i/u)(\partial / \partial q_{\nu}){\rm exp}(iu q\cdot x)$ so
that  the calculation can still be done in  momentum space.

\section{Numerical Results}\label{sec:numericalresult}
\label{sec:numericalResult}
In this section, we first give the numerical results for the  amplitudes of the
$\Xi_{cc}^{++}\to\Xi_{c}^{+(\prime)}\pi^{+}$ decays, namely the two constants $A_{\rm WE}^{(\prime)i}$ and
$B_{\rm WE}^{(\prime)i}$ in Eq.~(\ref{eq:parMatrix0WE}). 
In this work, we use the $\overline{\rm MS}$ masses for the quarks, $m_c(\mu)=1.27$~GeV and
$m_s(\mu)=0.103$~GeV with $\mu= 1.27$~GeV \cite{ParticleDataGroup:2020ssz}. The masses of the $u$
quark and the pion are neglected. The masses and decay constants of the charm baryons with positive
or negative parity are listed in Table~\ref{baryonMass}, where the decay constants are defined as
in Eq.~(\ref{XicDecayConst}).
\begin{table}
  \caption{The masses and decay constants of the charmed baryons with positive or negative parity.
    The decay constants are defined as in Eq.~(\ref{XicDecayConst}).}\label{baryonMass}
\begin{tabular}{|c|c|c|c|c|c|c|}
\hline 
 Baryon & $\Xi_{cc} (\frac{1}{2}^+)$ &  $\Xi_{cc}  (\frac{1}{2}^-)$ &  $\Xi_{c}  (\frac{1}{2}^+)$  &  $\Xi_{c}  (\frac{1}{2}^-)$  &  $\Xi_{c}^{\prime}  (\frac{1}{2}^+)$  & $\Xi_{c}^{\prime}  (\frac{1}{2}^-)$ \tabularnewline
\hline 
Mass ${\rm [GeV]}$ & $3.62$  \cite{ParticleDataGroup:2018ovx} & $3.77$ \cite{Wang:2010it} & $2.47$  \cite{ParticleDataGroup:2018ovx} & $2.79$ \cite{Roberts:2007ni} & $2.58$ \cite{ParticleDataGroup:2018ovx} & $2.87$\cite{Wang:2010it} \tabularnewline
\hline 
$\lambda$ ${\rm [GeV^3]}$ & $0.109$  \cite{Hu:2017dzi,1902.01092} & $0.159$\cite{Wang:2010it} & $0.038$  \cite{Hu:2017dzi,1902.01092} & $0.042$ \cite{Wang:2010fq} & $0.076$ \cite{Hu:2017dzi,1902.01092} & $0.084$ \cite{Wang:2010it} \tabularnewline
\hline 
\end{tabular}
\end{table}

On the other hand, the LCSR contains two kinds of extra parameters, the thresholds $s_{\Xi_{cc}},
s_{\Xi_c}, s_{\Xi_c^{\prime}}$ and the Borel parameters $T^2, T^{\prime 2}$. For the threshold parameters,
we have argued that they are process-independent and will be taken from Ref.~\cite{Khodjamirian:2000mi},
$s_{\Xi_{cc}}=(4.1\pm 0.1)^2$~GeV$^2$, $s_{\Xi_{c}}=(3.2\pm 0.1)^2$~GeV$^2$ and
$s_{\Xi_{c}^{\prime}}=(3.3\pm 0.1)^2$~GeV$^2$. These are about $0.5^2$~GeV$^2$ larger than 
the squared masses  of the corresponding baryons. Since this difference to the squared mass
is only an empirical value  proposed in Ref.~\cite{Wang:2012kw}, 
we will consider the uncertainty of the thresholds when evaluating the error of numerical results.

\begin{figure}
\begin{center}
\includegraphics[width=0.45\columnwidth]{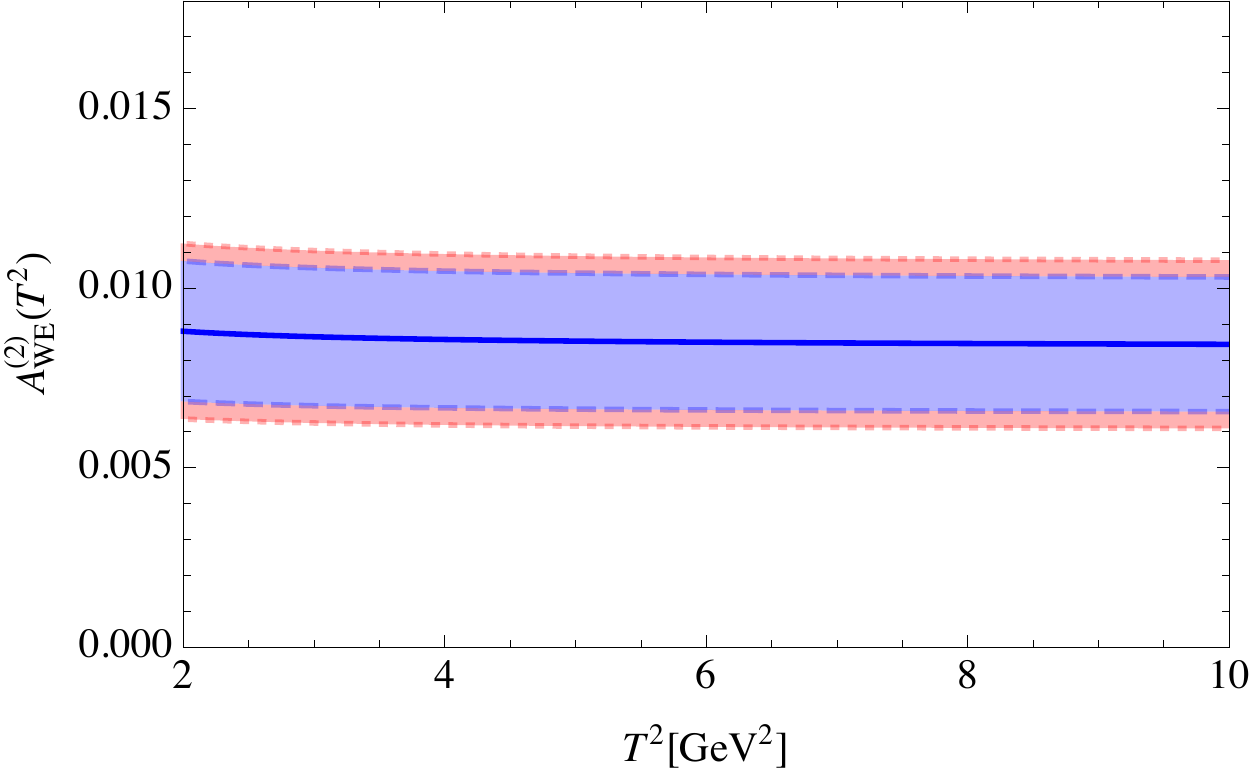} 
\includegraphics[width=0.45\columnwidth]{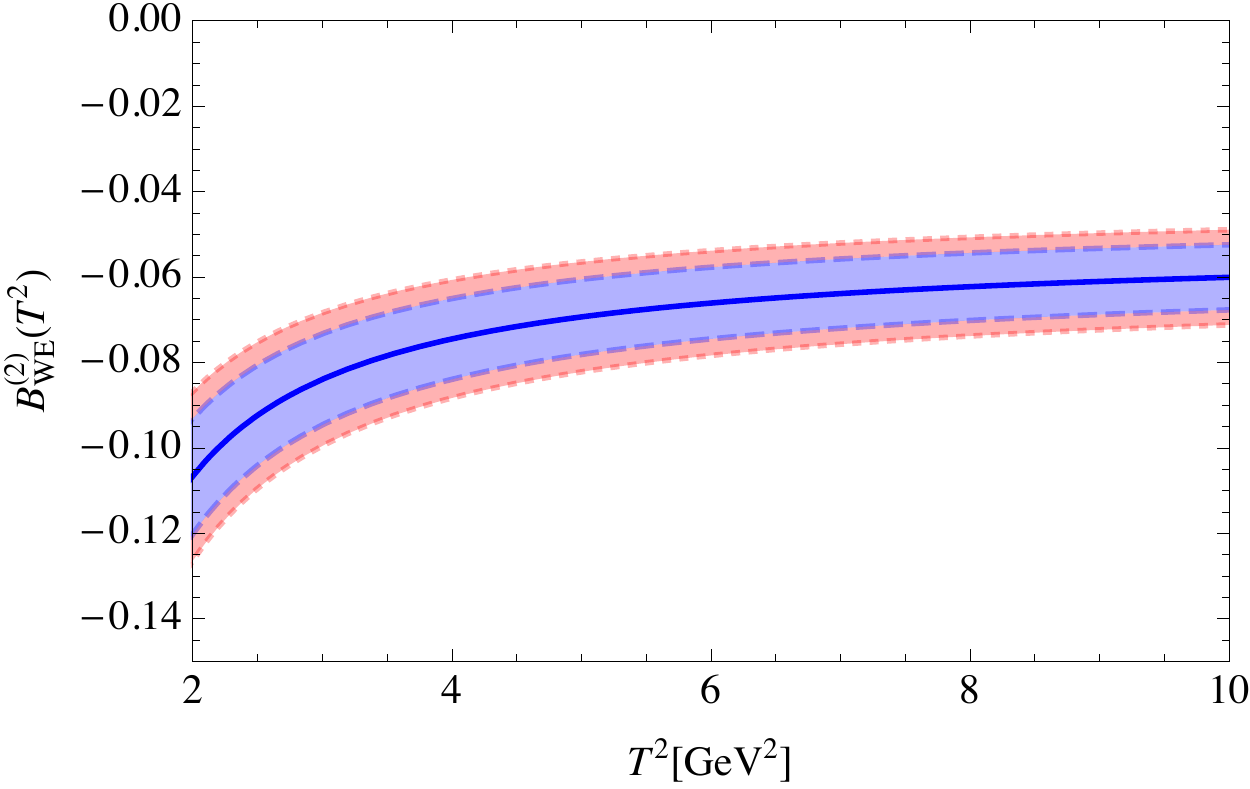} 
\includegraphics[width=0.45\columnwidth]{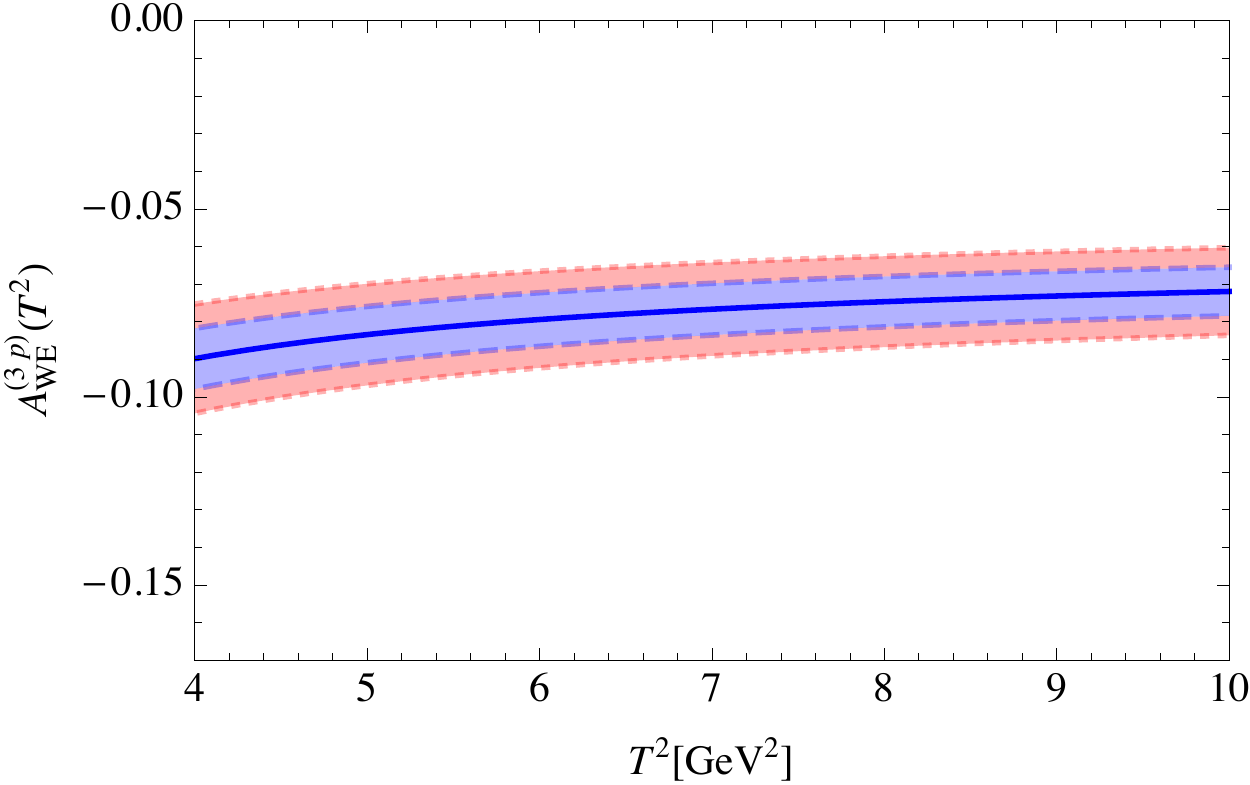} 
\includegraphics[width=0.45\columnwidth]{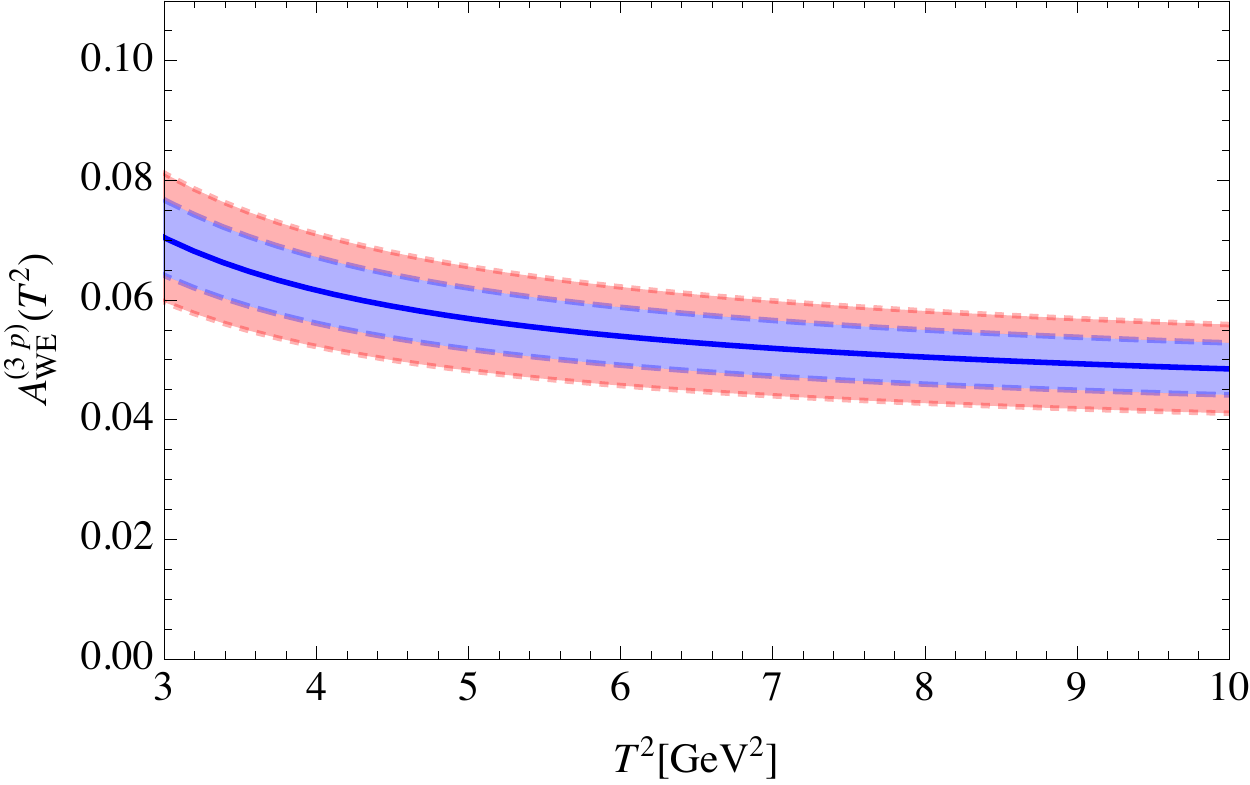} 
\includegraphics[width=0.45\columnwidth]{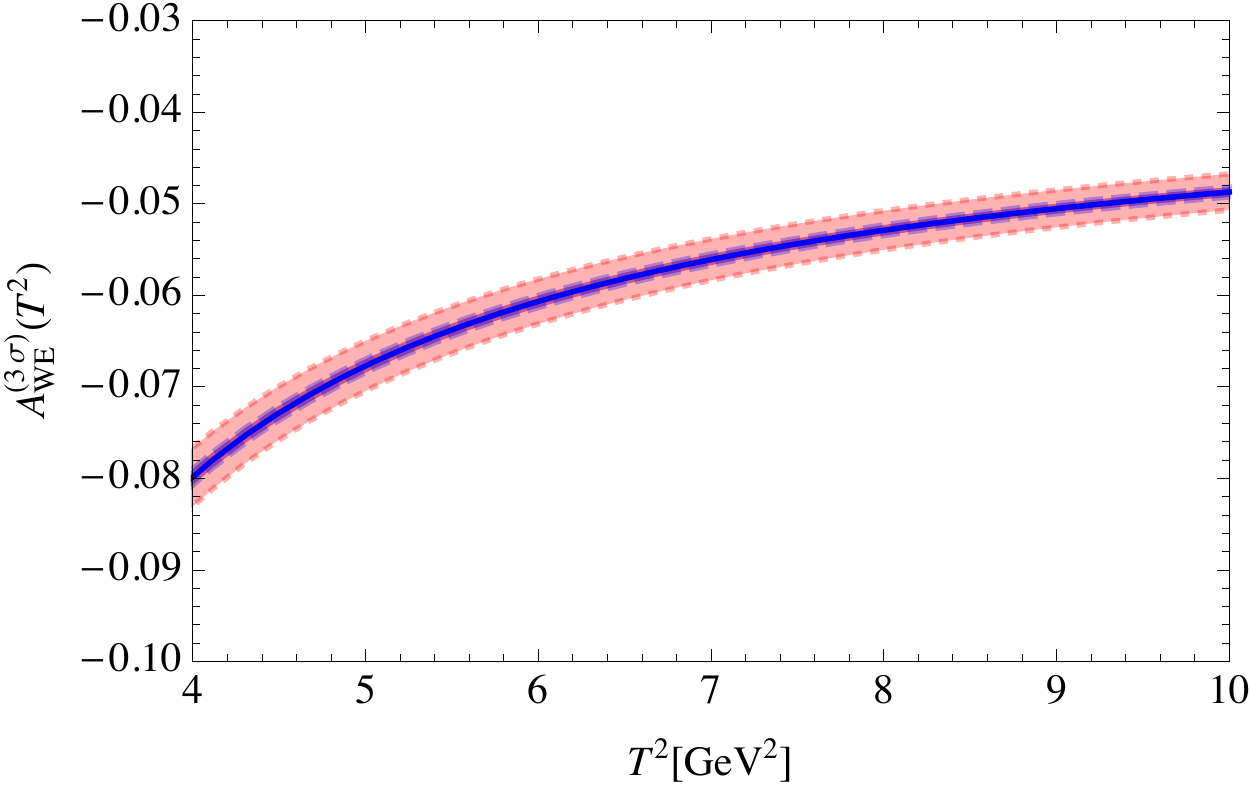} 
\includegraphics[width=0.45\columnwidth]{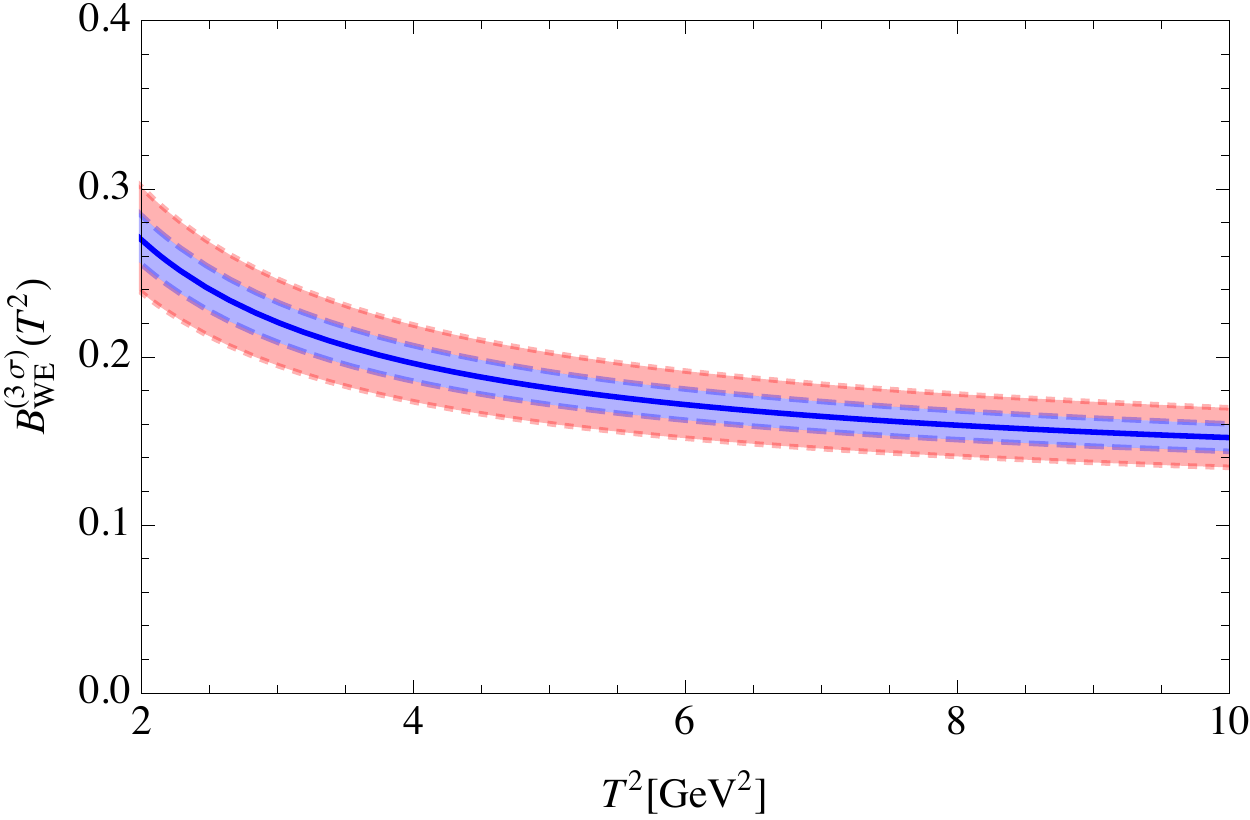} 
\caption{The Borel parameter dependence of the $\Xi_{cc}^{++}\to\Xi_{c}^{+}\pi^{+}$ decay amplitudes $A^{(2)}(B^{(2)}),A^{(3p)}(B^{(3p)})$ and $A^{(3\sigma)}(B^{(3\sigma)})$  from the twist-2, twist-3p and twist-3$\sigma$ LCDAs respectively. In each diagram, the blue band denotes the uncertainty from the error of the threshold: $s_{\Xi_{cc}}=(4.1\pm 0.1)^2$ GeV$^2$ and $s_{\Xi_{c}}=(3.2\pm 0.1)^2$ GeV$^2$. The upper and lower red bands denote the uncertainty from the error of Monte-Carlo integrations.}
\label{fig:ABvsT2} 
\end{center}
\end{figure}
\begin{figure}
\begin{center}
\includegraphics[width=0.45\columnwidth]{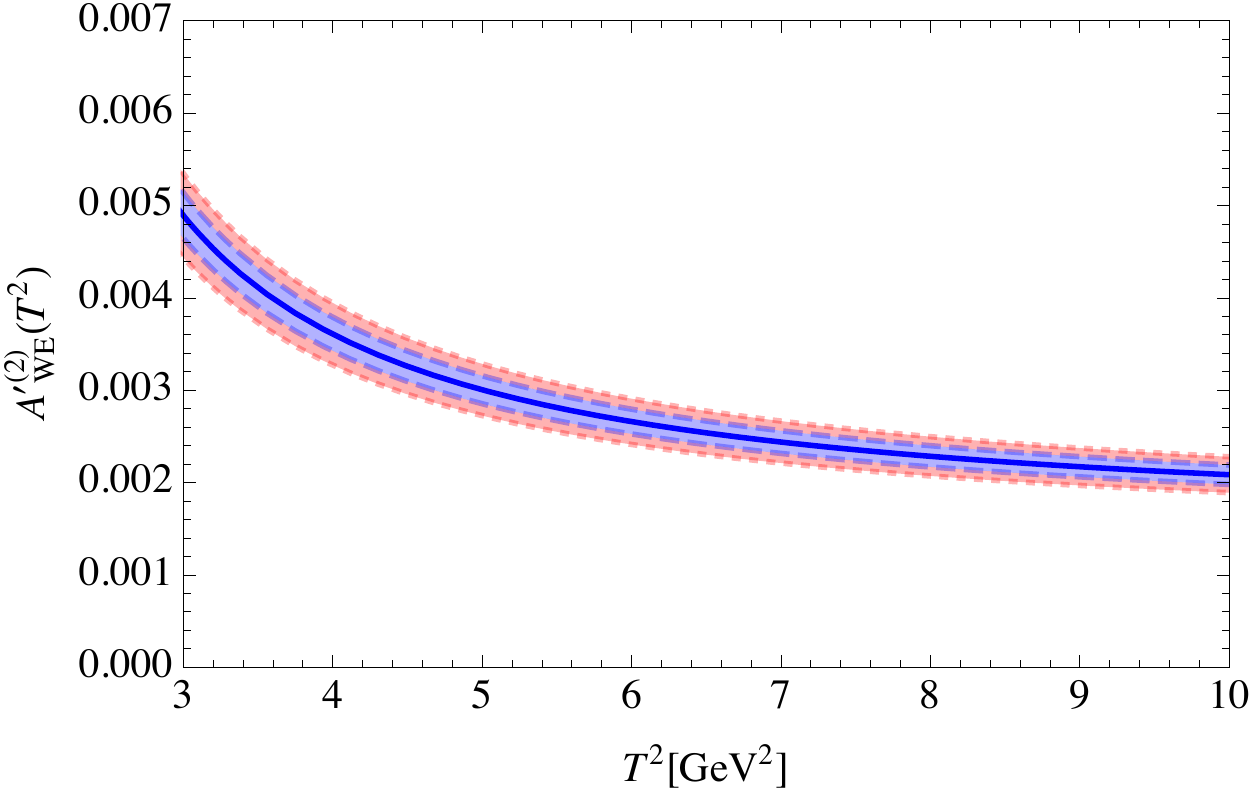} 
\includegraphics[width=0.45\columnwidth]{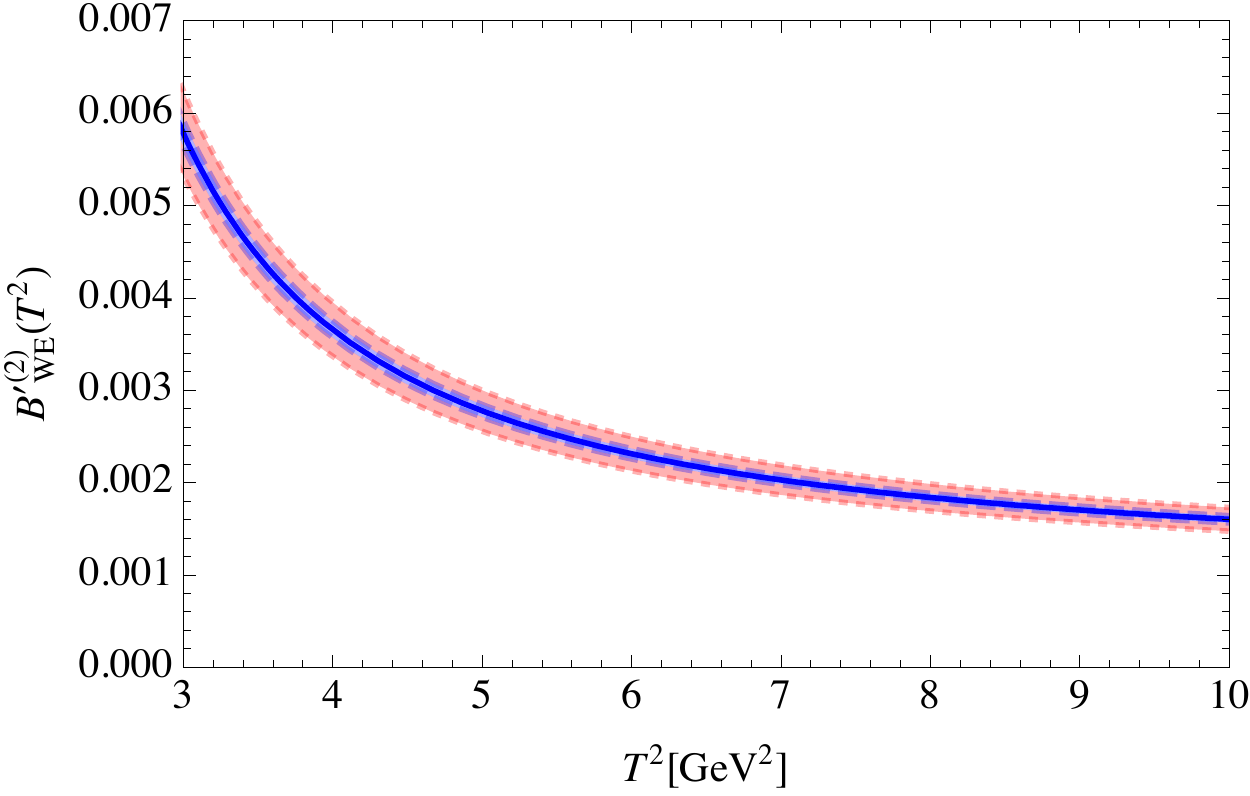} 
\includegraphics[width=0.45\columnwidth]{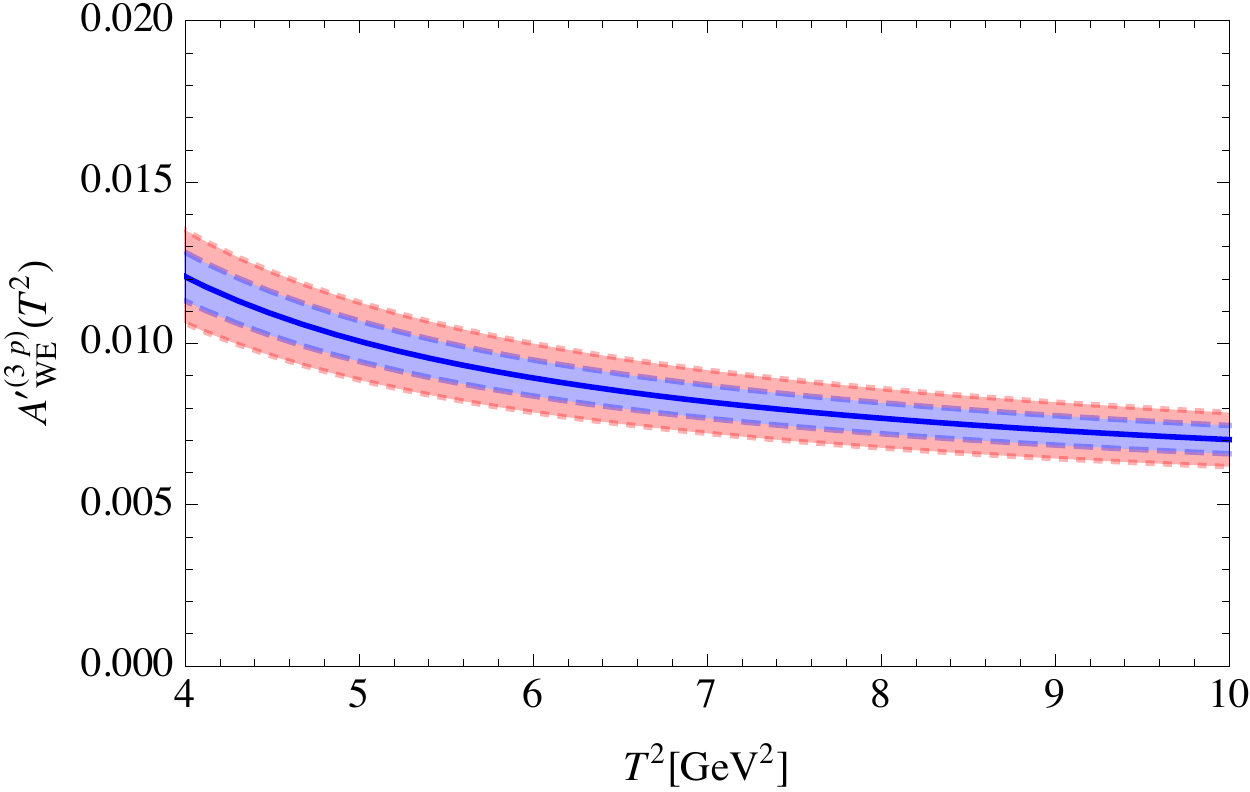} 
\includegraphics[width=0.45\columnwidth]{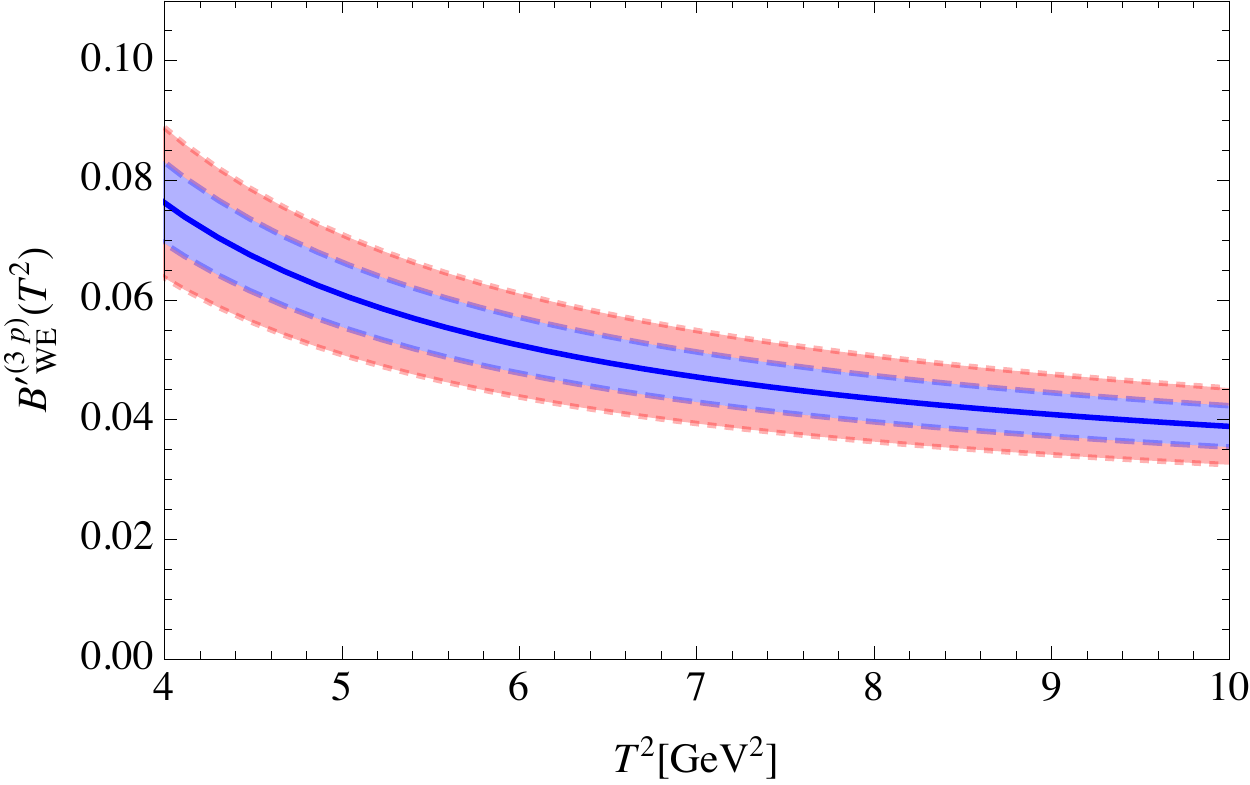} 
\includegraphics[width=0.45\columnwidth]{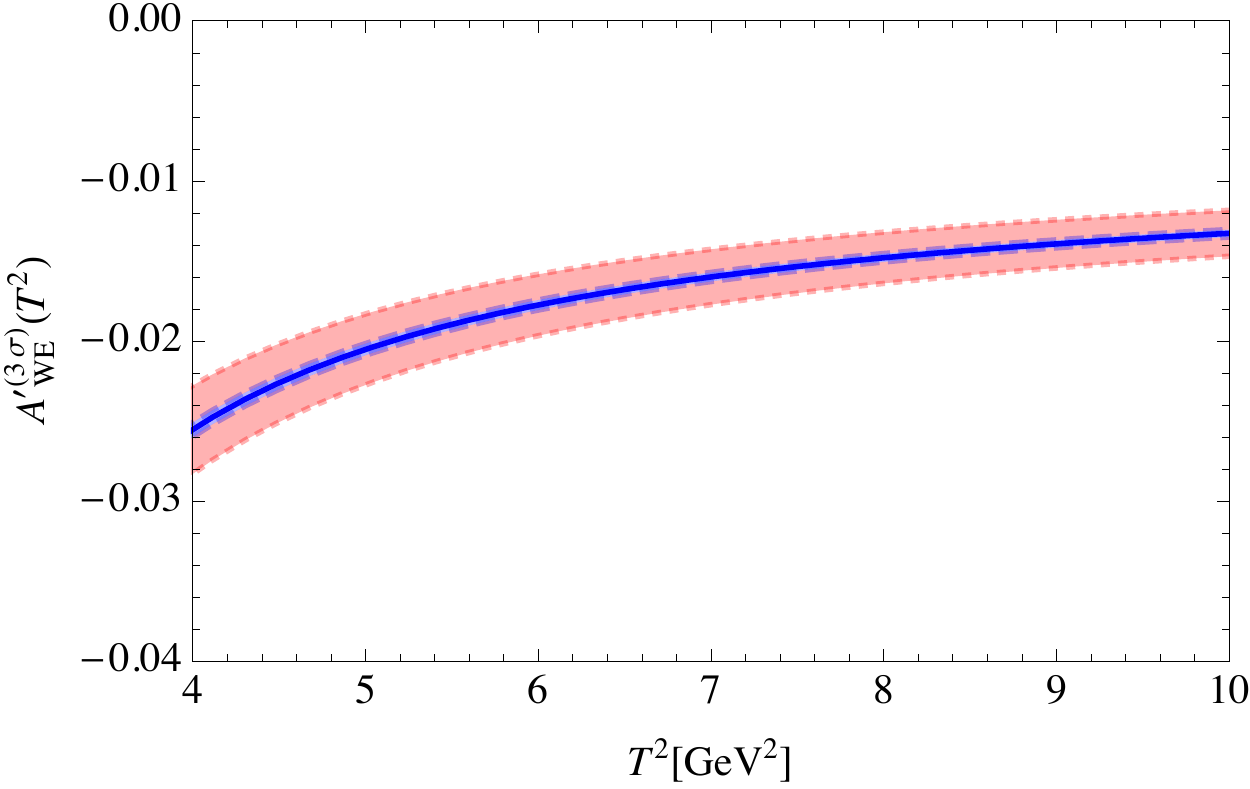} 
\includegraphics[width=0.45\columnwidth]{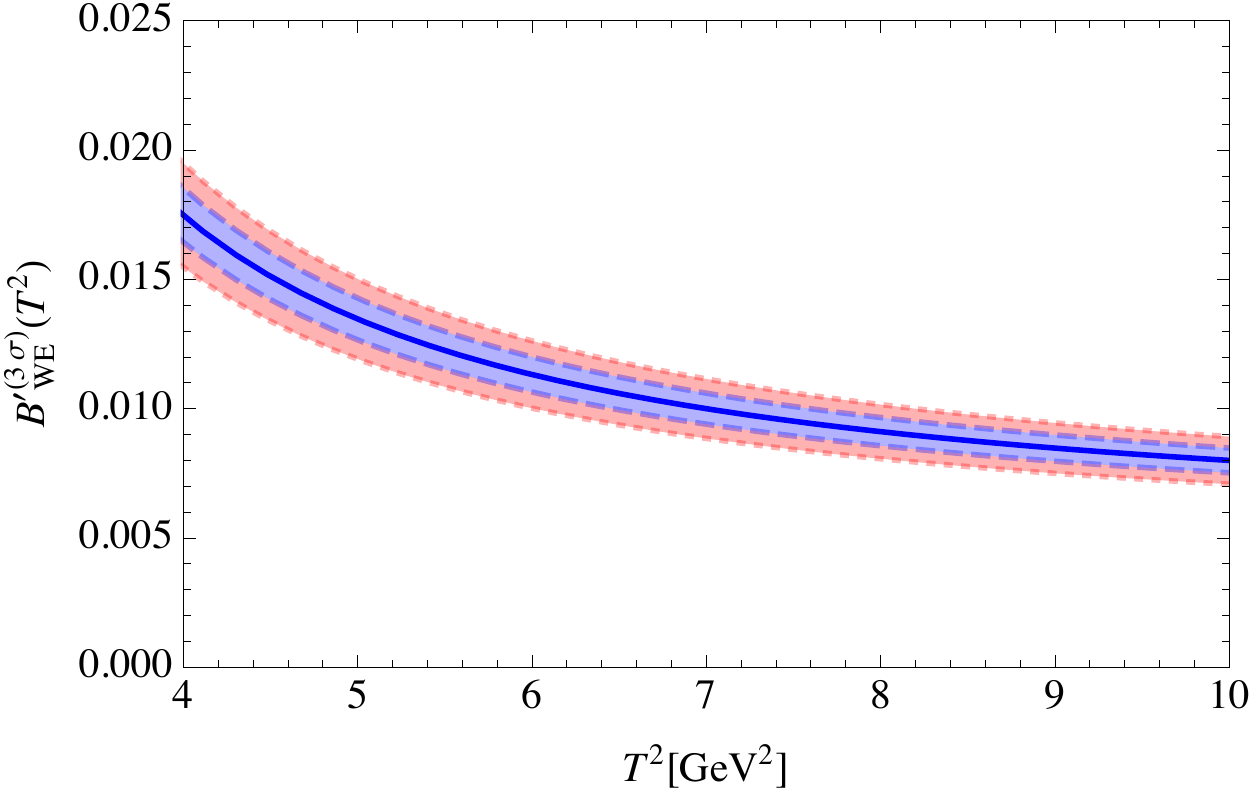} 
\caption{The Borel parameter dependence of the $\Xi_{cc}^{++}\to\Xi_{c}^{+\prime}\pi^{+}$ decay amplitudes $A^{\prime(2)}(B^{\prime(2)}),A^{\prime(3p)}(B^{\prime(3p)})$ and $A^{\prime(3\sigma)}(B^{\prime(3\sigma)})$ from the twist-2, twist-3p and twist-3$\sigma$ LCDAs respectively. In each diagram, the blue band denotes the uncertainty from the error of the threshold: $s_{\Xi_{cc}}=(4.1\pm 0.1)^2$ GeV$^2$ and $s_{\Xi_{c}^{\prime}}=(3.3\pm 0.1)^2$ GeV$^2$. The upper and lower red bands denote the uncertainty from the error of Monte-Carlo integrations.}
\label{fig:ABprimevsT2} 
\end{center}
\end{figure}
What concerns the Borel parameters, to simplify the problem we apply the following equation
to relate the two Borel parameters corresponding to the $s$ and $s^{\prime}$ channels \cite{Ball:1991bs}:
\begin{align}
\frac{T^{2}}{T^{\prime 2}} \approx \frac{M_{1}^{2}-m_{1}^{2}}{M_{2}^{2}-m_{2}}\equiv\frac{1}{\kappa},
\end{align}
where $M_{1(2)}$ is the mass of the initial (final) baryon and $m_{1(2)}$ is the mass of the quark before
(after) the weak decay. In Fig.~\ref{fig:ABvsT2} and Fig.~\ref{fig:ABprimevsT2} we present
the Borel parameter dependence of the $\Xi_{cc}^{++}\to\Xi_{c}^{+\prime}\pi^{+}$ decay amplitudes
$A^{(\prime)(2)}(B^{(\prime)(2)}),A^{(\prime)(3p)}(B^{(\prime)(3p)})$ and $A^{(\prime)(3\sigma)}(B^{(\prime)(3\sigma)})$
contributed from the twist-2, twist-3p and twist-3$\sigma$ LCDAs, respectively. In each diagram,
the blue band denotes the uncertainty from the error of the threshold. The upper and lower
red bands denote the uncertainty from the error of the Monte Carlo integrations. 
Generally, the window of Borel parameters is chosen to satisfy three requirements. The first one
is that they must be small enough so that the contribution from the continuous spectrum can be
suppressed, which determines their upper bound. The second one is that they must be large
enough to ensure the OPE convergence, which determines their lower bound. The last one is
that the result must be stable in this window. To determine the upper bound of the Borel
parameter, we require that the pole contribution must be larger than the continuous
spectrum contribution, namely:
\begin{align}
\frac{\int_{(m_{c}+m_{s})^{2}}^{s_{\Xi_{c}}}ds^{\prime}\int_{4m_{c}^{2}}^{s_{\Xi_{cc}}}ds\ e^{-s^{\prime}/\kappa T^{2}}e^{-s/T^{2}}{\rm Im}^{2}\Pi_{QCD}^{{\cal O}_{i}}(s^{\prime},s,P^{2})}{\int_{(m_{c}+m_{s})^{2}}^{\infty}ds^{\prime}\int_{4m_{c}^{2}}^{\infty}ds\ e^{-s^{\prime}/\kappa T^{2}}e^{-s/T^{2}}{\rm Im}^{2}\Pi_{QCD}^{{\cal O}_{i}}(s^{\prime},s,P^{2})}>0.5.
\end{align}
The numerator represents the pole contribution, which is just the integral
on the right-hand side of Eq.~(\ref{eq:sumruleEq}). The denominator is the same integral
but the upper limits of $s$ and $s^{\prime}$ are extended to infinity, so that it contains both
pole and continuous spectrum contributions. This requirement shows that the upper bound of
the Borel parameter is around $T^2=7$~GeV$^2$ for $\Xi_{cc}^{++}\to\Xi_{c}^{+}\pi^{+}$ and
$T^2=6$~GeV$^2$ for $\Xi_{cc}^{++}\to\Xi_{c}^{+\prime}\pi^{+}$.

The lower bounds of the Borel parameters are determined in principle by the ratio between
the contribution from the leading order and next-to-leading order QCD corrections to the perturbative
kernel of OPE. However, in this work we have only considered the leading order contribution
so that this method cannot be used. From Fig.~\ref{fig:ABvsT2} and Fig.~\ref{fig:ABprimevsT2},
it can be seen that the upper bounds given above are in a relatively stable region.
Therefore, although we cannot determine the lower bound quantitatively, we can take 
a range below the upper bound of the Borel parameter. Here, we set the window as
$6<T^2<8$~GeV$^2$ for $\Xi_{cc}^{++}\to\Xi_{c}^{+}\pi^{+}$ and $5<T^2<7$~GeV$^2$ for
$\Xi_{cc}^{++}\to\Xi_{c}^{+\prime}\pi^{+}$. The amplitudes and the corresponding errors from the
uncertainties of $s_{\Xi_{cc}}, s_{\Xi_{c}^{\prime}}$ and $T^2$ are listed in Table~\ref{Tab:decayAmps}. 
Note  that most of the contributions from the twist-3 LCDAs are larger than that from twist-2.
This feature is common in the LCSR studies on the heavy-to-light decays. 
For example, in the LCSR calculation of the $B\to\pi$ form factors \cite{Duplancic:2008ix,Rusov:2017chr},
the contribution of  twist-3 is generally of the same order or even larger than that of twist-2.
The highly suppressed contributions should come from the LCDAs of twist-4 or higher.

\begin{table}
  \caption{Decay amplitudes of $\Xi_{cc}^{++}\to\Xi_{c}^{+(\prime)}\pi^{+}$ from the W-exchange contribution.
    The Borel parameters are set in the region $6<T^2<8$~GeV$^2$ for $\Xi_{cc}^{++}\to\Xi_{c}^{+}\pi^{+}$
    and $5<T^2<7$~GeV$^2$ for $\Xi_{cc}^{++}\to\Xi_{c}^{+\prime}\pi^{+}$.}
\label{Tab:decayAmps}
\begin{tabular}{|c|cccc|}
\hline 
$\Xi_{cc}^{++}\to\Xi_{c}^{+}\pi^{+}$ & Twist-$2$ & Twist-$3p$ & Twist-$3\sigma$ & Total\tabularnewline
\hline 
$A_{{\rm WE}}$ & $0.0084\pm0.0024$ & $-0.077\pm0.01$ & $-0.056\pm0.002$ & $-0.124\pm0.011$\tabularnewline
$B_{{\rm WE}}$ & $-0.064\pm0.01$ & $0.052\pm0.01$ & $0.165\pm0.025$ & $0.153\pm0.029$\tabularnewline
\hline 
\hline 
$\Xi_{cc}^{++}\to\Xi_{c}^{+\prime}\pi^{+}$ & Twist-$2$ & Twist-$3p$ & Twist-$3\sigma$ & Total\tabularnewline
\hline 
$A_{{\rm WE}}^{\prime}$ & $0.0027\pm0.0005$ & $0.0089\pm0.002$ & $-0.018\pm0.0003$ & $-0.0062\pm0.002$\tabularnewline
$B_{{\rm WE}}^{\prime}$ & $0.0023\pm0.0006$ & $0.052\pm0.016$ & $0.011\pm0.003$ & $0.066\pm0.016$\tabularnewline
\hline 
\end{tabular}
\end{table}

\begin{table}
  \caption{Comparison of the decay amplitudes of $\Xi_{cc}^{++}\to\Xi_{c}^{+(\prime)}\pi^{+}$ from this work
    with those from the literature.  All the amplitudes below are in unit $10^{-2} G_F$~GeV$^2$.
    Here we list the W-exchange amplitudes from the pole model (PM)
    \cite{Cheng:2020wmk,Sharma:2017txj,Dhir:2018twm}
    and three-loop quark model (3LQM) \cite{Gutsche:2018msz}. We also list the factorizable amplitudes
    from QCD sum rules (QCDSR) \cite{Shi:2019hbf}, light-front quark model (LFQM)\cite{Cheng:2020wmk},
    nonrelativistic quark model (NRQM) \cite{Sharma:2017txj,Dhir:2018twm} and heavy quark effective
    theory (HQET) \cite{Sharma:2017txj,Dhir:2018twm}. The notation LFQM + PM for example in
    the first column means the factorizable amplitude is from LFQM while the W-exchange
    amplitude is from PM, otherwise both of them are from the same theoretical approach. }
\label{Tab:decayAmpsCompar}
\begin{tabular}{|c|cccccc|}
\hline 
$\Xi_{cc}^{++}\to\Xi_{c}^{+}\pi^{+}$ & $A^{{\rm fac}}$ & $A^{{\rm nf}}$ & $A^{{\rm tot}}$ & $B^{{\rm fac}}$ & $B^{{\rm nf}}$ & $B^{{\rm tot}}$\tabularnewline
\hline 
This work & $--$ & $-16.67\pm1.41$ & $--$ & $--$ & $20.47\pm3.89$ & $--$\tabularnewline
QCDSR \cite{Shi:2019hbf} & $-8.74\pm2.91$ & $--$ & $--$ & $16.76\pm5.36$ & $--$ &$--$ \tabularnewline
LFQM + PM \cite{Cheng:2020wmk} & $7.40$ & $-10.79$ & $-3.83$ & $15.06$  & $-18.91$ & $3.85$\tabularnewline
3LQM \cite{Gutsche:2018msz} & $-8.13$ & $10.50$ & $3.37$ & $-12.97$ & $18.53$ & $5.56$\tabularnewline
NRQM + PM \cite{Sharma:2017txj,Dhir:2018twm} & $7.38$ & $0$ & $7.38$ & $16.77$ & $24.95$ & $41.72$\tabularnewline
HQET + PM \cite{Sharma:2017txj,Dhir:2018twm} & $9.52$ & $0$ & $9.52$ & $19.45$ & $24.95$ & $44.40$\tabularnewline
\hline 
\hline 
$\Xi_{cc}^{++}\to\Xi_{c}^{+\prime}\pi^{+}$ & $A^{\prime{\rm fac}}$ & $A^{\prime{\rm nf}}$ & $A^{\prime{\rm tot}}$ & $B^{\prime{\rm fac}}$ & $B^{{\rm \text{\ensuremath{\prime}}nf}}$ & $B^{{\rm \prime tot}}$\tabularnewline
\hline 
This Work & $--$ & $-0.83\pm0.28$ & $--$ & $--$ & $8.86\pm2.16$ & $--$ \tabularnewline
QCDSR \cite{Shi:2019hbf} & $-3.55\pm0.68$ & $--$ & $--$ & $34.13\pm11.6$ & $--$ &$--$ \tabularnewline
LFQM + PM \cite{Cheng:2020wmk} & $4.49$ & $-0.04$ & $4.45$ & $48.50$ & $-0.06$ & $48.44$\tabularnewline
3LCQM \cite{Gutsche:2018msz} & $-4.34$ & $-0.11$ & $-4.45$ & $-37.59$ & $-1.37$ & $-38.96$\tabularnewline
NRQM + PM \cite{Sharma:2017txj,Dhir:2018twm} & $4.29$ & $0$ & $4.29$ & $53.65$ & $0$ & $53.65$\tabularnewline
HQET + PM \cite{Sharma:2017txj,Dhir:2018twm} & $5.10$ & $0$ & $5.10$ & $62.37$ & $0$ & $62.37$\tabularnewline
\hline 
\end{tabular}
\end{table}
Table~\ref{Tab:decayAmpsCompar} shows the comparison of our results with those from the literature.
Here, we have unified the definition of the amplitudes from all of  these works. The amplitudes
presented in the table are the parameters of the matrix element induced by the effective
Hamiltonian from Eq.~(\ref{eq:effHamil}) instead of ${\cal O}_{1,2}$:
\begin{align}
  \langle\Xi_{c}^{+(\prime)}(p-q)\pi^{+}(q)|{\cal H}_{\rm eff}(0)|\Xi_{cc}^{++}(p)\rangle_{\rm fac, nf}
  =i\  \bar{u}(p-q)\big[A^{(\prime) \rm fac, nf}+B^{(\prime)\rm fac, nf}\gamma_{5}\big]u(p),\label{eq:parMatrix0}
\end{align}
where ``fac'' means the factorizable or W-emission contribution while ``nf'' denotes the non-factorizable or
equivalently the W-exchange contribution, and $(A,B)^{(\prime) \rm tot}=(A,B)^{(\prime) \rm fac}
+(A,B)^{(\prime) \rm nf}$.  All the amplitudes in Table~\ref{Tab:decayAmpsCompar} are in units of
$10^{-2} G_F$~GeV$^2$. Here we have listed the W-exchange amplitudes from the pole model (PM)
\cite{Cheng:2020wmk,Sharma:2017txj,Dhir:2018twm} and three-loop constituent quark model (3LCQM)
\cite{Gutsche:2018msz}. We have also listed the corresponding  factorizable amplitudes from
QCDSR \cite{Shi:2019hbf}, LFQM \cite{Cheng:2020wmk}, nonrelativistic quark model (NRQM)
\cite{Sharma:2017txj,Dhir:2018twm} and heavy quark effective theory (HQET)
\cite{Sharma:2017txj,Dhir:2018twm}. The notation LFQM + PM for example in the first column
means that the factorizable amplitude is from LFQM while the W-exchange amplitude is from PM.
On the other hand, 3LCQM alone means that the factorizable and non-factorizable amplitudes
are both from 3LCQM. 
From the comparison in Table~\ref{Tab:decayAmpsCompar}, one can find that our results for the
W-exchange amplitudes of $\Xi_{cc}^{++}\to\Xi_{c}^{+}\pi^{+}$ are consistent with most of the
results from literatures. However, in terms of the W-exchange amplitudes of
$\Xi_{cc}^{++}\to\Xi_{c}^{+\prime}\pi^{+}$, our results are much larger than those from  the literature,
which provides a possibility to explain the anomaly of ${\cal B}^{\prime}/{\cal B}$. 
Two issues deserve further discussion::
\begin{itemize}
\item We have only used the two-particle LCDAs of the pion for the calculation up to the leading order.
Although the contribution from the higher-twist LCDAs or QCD loop corrections is expected to be suppressed,
in principle they are still necessary for improving the accuracy of the W-exchange contribution, which will be
included in a future study.

\item From Table~\ref{Tab:decayAmpsCompar}, we see that the various predictions on the factorizable
contribution from the literature are not totally consistent with each other. These works assume the
naive factorization which is based on color transparency \cite{Bjorken:1988kk}. It states that in the
bottom hadron decays, the $b$ quark is heavy enough so that the emitted light meson flies quickly and
decouples from other hadrons before it is caught up by the soft gluons. However, since the charm quark
is lighter, in the charmed hadron decays the effect of soft gluon exchange may be important. Therefore,
such effects need to be worked out in the future.
\end{itemize}

\begin{table}
  \caption{Comparison of the decay branching fractions of $\Xi_{cc}^{++}\to\Xi_{c}^{+(\prime)}\pi^{+}$
    from this work with those from the literature. The lifetime of  the $\Xi_{cc}^{++}$ is chosen
    as $\tau(\Xi_{cc}^{++})=2.56\times 10^{-13} s$ \cite{Cheng:2020wmk}. In the first five lines
    the branching fractions are evaluated by the non-factorizable amplitudes from this work
    and the factorizable amplitudes from the literature. The last four lines present the branching
    fractions with both these two kinds of amplitudes evaluated in the literature. In the
    last column we list the values of ${\cal B}^{\prime}/{\cal B}$ from all the theoretical methods
    mentioned in Table~\ref{Tab:decayAmpsCompar}.}
\label{Tab:BrFCompa}
\resizebox{\textwidth}{25mm}{
\begin{tabular}{|c|c|c|c|c|c|c|c|}
\hline 
Method & $A^{{\rm tot}}$ & $B^{{\rm tot}}$ & ${\cal B}(\Xi_{cc}^{++}\to\Xi_{c}^{+}\pi^{+})$  & $A^{\prime{\rm tot}}$ & $B^{\prime{\rm tot}}$ & ${\cal B}(\Xi_{cc}^{++}\to\Xi_{c}^{+\prime}\pi^{+})$ & ${\cal B}^{\prime}/{\cal B}$\tabularnewline
\hline 
QCDSR+LCSR & $-25.4\pm4.32$ & $37.23\pm9.25$ & $40\pm14$ \% & $-4.38\pm0.96$ & $42.99\pm13.76$ & $3.91\text{\ensuremath{\pm}2.5}$ \% & $0.098\pm0.14$\tabularnewline
LFQM+LCSR & $-9.27\pm1.41$ & $35.53\pm3.89$ & $7.54\pm2.22$ \% & $3.66\pm0.28$ & $57.36\pm2.16$ & $5.83\pm0.5$ \% & $0.77\pm0.42$\tabularnewline
3LCQM+LCSR & $-24.8\pm1.41$ & $7.5\pm3.89$ & $35.55\pm4.29$ \% & $-5.17\pm0.28$ & $-28.73\pm2.16$ & $2.75\pm0.35$ \% & $0.08\pm0.02$\tabularnewline
NRQM+LCSR & $-9.29\pm1.41$ & $37.24\pm3.89$ & $7.82\pm2.25$ \% & $3.46\pm0.28$ & $62.51\pm2.16$ & $6.70\pm0.54$ \% & $0.85\pm0.44$\tabularnewline
HQET+LCSR & $-7.18\pm1.41$ & $39.92\pm3.89$ & $6.22\pm1.94$ \% & $4.27\pm0.28$ & $71.23\pm2.16$ & $8.85\pm0.62$ \% & $1.42\pm0.78$\tabularnewline
\hline 
\hline 
LFQM+PM & $-3.83$ & 3.85 & $0.69$ \% & $4.45$ & $48.44$ & $4.65$ \%  & $6.74$\tabularnewline
3LCQM & $3.37$ & $5.56$ & $0.71$ \% & $-4.45$ & $-38.96$ & $3.39$ \% & $4.77$\tabularnewline
HQET+PM & $7.38$ & $41.72$ & $6.64$ \% & $4.29$ & $53.65$ & $5.39$ \% & $0.81$\tabularnewline
NRQM+PM & 9.52 & 44.40 & $9.19$ \% & $5.1$ & $62.37$ & $7.34$ \% & $0.8$\tabularnewline
\hline 
\end{tabular}}
\end{table}
Now we calculate the decay branching fractions by combining each factorizable amplitude
from the literature and the non-factorizable amplitudes from this work, and make a comparison
among them. The decay width is expressed as
\begin{align}
\Gamma^{(\prime)}=\frac{p_{c}}{8 \pi}\left[\frac{\big(m_{\Xi_{cc}}+m_{\Xi_{c}^{(\prime)}}\big)^{2}-m_{\pi}^{2}}{m_{\Xi_{cc}}^{2}}|A^{\rm tot(\prime)}|^{2}+\frac{\big(m_{\Xi_{cc}}-m_{\Xi_{c}^{(\prime)}}\big)^{2}-m_{\pi}^{2}}{m_{\Xi_{cc}}^{2}}|B^{\rm tot(\prime)}|^{2}\right],
\end{align}
where $p_c$ is the magnitude of the pion three-momentum in the rest-frame of the $\Xi_{cc}$.
The lifetime of the $\Xi_{cc}^{++}$ is chosen as $\tau(\Xi_{cc}^{++})=2.56\times 10^{-13} s$,
the Wilson coefficients are chosen as $C_1=1.35$ and $C_2=-0.64$ \cite{Cheng:2020wmk},
$G_{F}=1.166\times 10^{-5}$ GeV$^{-2}$ and $V_{cs}=0.975,\ V_{ud}=0.973$ \cite{ParticleDataGroup:2020ssz}.
The branching fractions are listed in Table~\ref{Tab:BrFCompa}. In the first five lines
the branching fractions are evaluated by the non-factorizable amplitudes from this work
and the factorizable amplitudes from the literature. The last four lines present the branching
fractions with both kinds of amplitudes being evaluated in the literature. In the last
column we list the values of ${\cal B}^{\prime}/{\cal B}$ from all the theoretical methods
mentioned above. We note that most of the calculated ${\cal B}^{\prime}/{\cal B}$ are much larger
or smaller than the experimental value $({\cal B}^{\prime}/{\cal B})_{\rm exp}=1.41\pm 0.17\pm 0.1$.
However, the value obtained  from HQET+LCSR leads to the fraction
\begin{equation}
({\cal B}^{\prime}/{\cal B})_{\rm HQET+LCSR}=1.42\pm0.78~,
\end{equation}
which agrees amazingly well with the
experimental value.
Although there exists non-negligible uncertainty for this theoretical result,
it still implies that $({\cal B}^{\prime}/{\cal B})_{\rm exp}=1.41>1$ can be understood and
realized theoretically without introducing any physics beyond the Standard Model.
The absolute branching fraction of $\Xi_{cc}^{++}\to\Xi_{c}^{+}\pi^{+}$
from HQET+LCSR is 
\begin{align}
{\cal B}(\Xi_{cc}^{++}\to\Xi_{c}^{+}\pi^{+})_{\rm HQET+LCSR}=6.22\pm1.94\  \%. 
\end{align}
Up to now, there is no experimental announcement on this absolute branching fraction. Instead, in Ref.~\cite{Cheng:2020wmk} the authors have given an evaluation of it by phenomenological  methods:
\begin{align}
{\cal B}(\Xi_{cc}^{++}\to\Xi_{c}^{+}\pi^{+}) =1.83\pm1.01\  \%,
\end{align}
which is consistent with our result within the uncertainties. For the details on this phenomenological evaluation we refer to the discussion around the Eq.~(38) in Ref.~\cite{Cheng:2020wmk}.  	
However,  due to the lack of experimental measurements all the theoretical predictions given above are still waiting to be tested by future experiments.

\section{Conclusion}
\label{sec:conclusion}
We have calculated the W-exchange contribution in the $\Xi_{cc}^{++}\to\Xi_{c}^{+(\prime)}\pi^{+}$ decay
with the use of the LCSR. The two-particle LCDAs of pion are used as the non-perturbative inputs
for the sum rules calculation, and the perturbative kernel is calculated at the leading order.
We obtain the relative decay branching fraction ${\cal B}^{\prime}/{\cal B}$ of $\Xi_{cc}^{++}\to
\Xi_{c}^{+(\prime)}\pi^{+}$ by combining our W-exchange amplitudes with the factorizable amplitudes
from various theoretical methods in the literature. We find that using the factorizable amplitudes
from  heavy quark effective theory, we obtain the ratio ${\cal B}^{\prime}/{\cal B}= 1.42\pm 0.78$,
which is consistent with the experimental value, $1.41\pm 0.17\pm 0.1$. The corresponding absolute
decay branching fraction is consistent with that evaluated in the literature, which should
be tested by future experiments.

\section*{Acknowledgements}
This work is supported in part by the NSFC and the Deutsche Forschungsgemeinschaft (DFG, German Research
Foundation) through the funds provided to the Sino-German Collaborative
Research Center TRR110 ``Symmetries and the Emergence of Structure in QCD''
(NSFC Grant  No. 12070131001, DFG Project-ID 196253076 - TRR 110). 
The work of UGM was supported in part by the Chinese
Academy of Sciences (CAS) President's International
Fellowship Initiative (PIFI) (Grant No. 2018DM0034)
and by VolkswagenStiftung (Grant No. 93562). The work of ZXZ is supported in
part by National Science Foundation of China
under Grant No. 12065020. The work of YX is supported in part by National Science Foundation of China
under Grant No. 12005294.


\begin{thebibliography}{10}
\bibitem{Gell-Mann:1964ewy}
M.~Gell-Mann,
Phys. Lett. \textbf{8}, 214-215 (1964)
doi:10.1016/S0031-9163(64)92001-3

\bibitem{Zweig:1964jf}
G.~Zweig,
CERN-TH-412.

\bibitem{DeRujula:1975qlm}
A.~De Rujula, H.~Georgi and S.~L.~Glashow,
Phys. Rev. D \textbf{12}, 147-162 (1975)
doi:10.1103/PhysRevD.12.147

\bibitem{Jaffe:1975us}
R.~L.~Jaffe and J.~E.~Kiskis,
Phys. Rev. D \textbf{13}, 1355 (1976)
doi:10.1103/PhysRevD.13.1355

\bibitem{Ponce:1978gk}
W.~Ponce,
Phys. Rev. D \textbf{19}, 2197 (1979)
doi:10.1103/PhysRevD.19.2197

\bibitem{Fleck:1988vm}
S.~Fleck, B.~Silvestre-Brac and J.~M.~Richard,
Phys. Rev. D \textbf{38}, 1519-1529 (1988)
doi:10.1103/PhysRevD.38.1519

\bibitem{LHCb:2017iph}
R.~Aaij \textit{et al.} [LHCb],
Phys. Rev. Lett. \textbf{119}, no.11, 112001 (2017)
doi:10.1103/PhysRevLett.119.112001
[arXiv:1707.01621 [hep-ex]].

\bibitem{Yu:2017zst}
F.~S.~Yu, H.~Y.~Jiang, R.~H.~Li, C.~D.~L\"u, W.~Wang and Z.~X.~Zhao,
Chin. Phys. C \textbf{42}, no.5, 051001 (2018)
doi:10.1088/1674-1137/42/5/051001
[arXiv:1703.09086 [hep-ph]].

\bibitem{LHCb:2018pcs}
R.~Aaij \textit{et al.} [LHCb],
Phys. Rev. Lett. \textbf{121}, no.16, 162002 (2018)
doi:10.1103/PhysRevLett.121.162002
[arXiv:1807.01919 [hep-ex]].

\bibitem{LHCb:2022rpd}
R.~Aaij \textit{et al.} [LHCb],
JHEP \textbf{05}, 038 (2022)
doi:10.1007/JHEP05(2022)038
[arXiv:2202.05648 [hep-ex]].

\bibitem{Bjorken:1988kk}
J.~D.~Bjorken,
Nucl. Phys. B Proc. Suppl. \textbf{11}, 325-341 (1989)
doi:10.1016/0920-5632(89)90019-4

\bibitem{Wirbel:1985ji}
M.~Wirbel, B.~Stech and M.~Bauer,
Z. Phys. C \textbf{29}, 637 (1985)
doi:10.1007/BF01560299

\bibitem{Bauer:1986bm}
M.~Bauer, B.~Stech and M.~Wirbel,
Z. Phys. C \textbf{34}, 103 (1987)
doi:10.1007/BF01561122

\bibitem{Shi:2019hbf}
Y.~J.~Shi, W.~Wang and Z.~X.~Zhao,
Eur. Phys. J. C \textbf{80}, no.6, 568 (2020)
doi:10.1140/epjc/s10052-020-8096-2
[arXiv:1902.01092 [hep-ph]].

\bibitem{Shi:2019fph}
Y.~J.~Shi, Y.~Xing and Z.~X.~Zhao,
Eur. Phys. J. C \textbf{79}, no.6, 501 (2019)
doi:10.1140/epjc/s10052-019-7014-y
[arXiv:1903.03921 [hep-ph]].

\bibitem{Hu:2019bqj}
X.~H.~Hu and Y.~J.~Shi,
Eur. Phys. J. C \textbf{80}, no.1, 56 (2020)
doi:10.1140/epjc/s10052-020-7635-1
[arXiv:1910.07909 [hep-ph]].

\bibitem{Hu:2022xzu}
X.~H.~Hu and Y.~J.~Shi,
[arXiv:2202.07540 [hep-ph]].

\bibitem{Aliev:2022tvs}
T.~M.~Aliev, S.~Bilmis and M.~Savci,
[arXiv:2206.08253 [hep-ph]].

\bibitem{Aliev:2022maw}
T.~M.~Aliev, S.~Bilmis and M.~Savci,
[arXiv:2205.14012 [hep-ph]].

\bibitem{Sharma:2017txj}
N.~Sharma and R.~Dhir,
Phys. Rev. D \textbf{96}, no.11, 113006 (2017)
doi:10.1103/PhysRevD.96.113006
[arXiv:1709.08217 [hep-ph]].

\bibitem{Gerasimov:2019jwp}
A.~S.~Gerasimov and A.~V.~Luchinsky,
Phys. Rev. D \textbf{100}, no.7, 073015 (2019)
doi:10.1103/PhysRevD.100.073015
[arXiv:1905.11740 [hep-ph]].

\bibitem{Shi:2020qde}
Y.~J.~Shi, W.~Wang, Z.~X.~Zhao and U.-G.~Mei\ss{}ner,
Eur. Phys. J. C \textbf{80}, no.5, 398 (2020)
doi:10.1140/epjc/s10052-020-7949-z
[arXiv:2002.02785 [hep-ph]].

\bibitem{Zhao:2018mrg}
Z.~X.~Zhao,
Eur. Phys. J. C \textbf{78}, no.9, 756 (2018)
doi:10.1140/epjc/s10052-018-6213-2
[arXiv:1805.10878 [hep-ph]].

\bibitem{Wang:2017mqp}
W.~Wang, F.~S.~Yu and Z.~X.~Zhao,
Eur. Phys. J. C \textbf{77}, no.11, 781 (2017)
doi:10.1140/epjc/s10052-017-5360-1
[arXiv:1707.02834 [hep-ph]].

\bibitem{Cheng:2020wmk}
H.~Y.~Cheng, G.~Meng, F.~Xu and J.~Zou,
Phys. Rev. D \textbf{101}, no.3, 034034 (2020)
doi:10.1103/PhysRevD.101.034034
[arXiv:2001.04553 [hep-ph]].

\bibitem{Ke:2019lcf}
H.~W.~Ke, F.~Lu, X.~H.~Liu and X.~Q.~Li,
Eur. Phys. J. C \textbf{80}, no.2, 140 (2020)
doi:10.1140/epjc/s10052-020-7699-y
[arXiv:1912.01435 [hep-ph]].

\bibitem{Ke:2022gxm}
H.~W.~Ke and X.~Q.~Li,
Phys. Rev. D \textbf{105}, no.9, 096011 (2022)
doi:10.1103/PhysRevD.105.096011
[arXiv:2203.10352 [hep-ph]].

\bibitem{Hu:2020mxk}
X.~H.~Hu, R.~H.~Li and Z.~P.~Xing,
Eur. Phys. J. C \textbf{80}, no.4, 320 (2020)
doi:10.1140/epjc/s10052-020-7851-8
[arXiv:2001.06375 [hep-ph]].

\bibitem{Gutsche:2018msz}
T.~Gutsche, M.~A.~Ivanov, J.~G.~K\"orner, V.~E.~Lyubovitskij and Z.~Tyulemissov,
Phys. Rev. D \textbf{99}, no.5, 056013 (2019)
doi:10.1103/PhysRevD.99.056013
[arXiv:1812.09212 [hep-ph]].

\bibitem{Gutsche:2019iac}
T.~Gutsche, M.~A.~Ivanov, J.~G.~K\"orner, V.~E.~Lyubovitskij and Z.~Tyulemissov,
Phys. Rev. D \textbf{100}, no.11, 114037 (2019)
doi:10.1103/PhysRevD.100.114037
[arXiv:1911.10785 [hep-ph]].

\bibitem{Dhir:2018twm}
R.~Dhir and N.~Sharma,
Eur. Phys. J. C \textbf{78}, no.9, 743 (2018)
doi:10.1140/epjc/s10052-018-6220-3

\bibitem{Wang:2017azm}
W.~Wang, Z.~P.~Xing and J.~Xu,
Eur. Phys. J. C \textbf{77}, no.11, 800 (2017)
doi:10.1140/epjc/s10052-017-5363-y
[arXiv:1707.06570 [hep-ph]].

\bibitem{Shi:2017dto}
Y.~J.~Shi, W.~Wang, Y.~Xing and J.~Xu,
Eur. Phys. J. C \textbf{78}, no.1, 56 (2018)
doi:10.1140/epjc/s10052-018-5532-7
[arXiv:1712.03830 [hep-ph]].

\bibitem{Li:2017ndo}
R.~H.~Li, C.~D.~L\"u, W.~Wang, F.~S.~Yu and Z.~T.~Zou,
Phys. Lett. B \textbf{767}, 232-235 (2017)
doi:10.1016/j.physletb.2017.02.003
[arXiv:1701.03284 [hep-ph]].

\bibitem{Balitsky:1989ry}
I.~I.~Balitsky, V.~M.~Braun and A.~V.~Kolesnichenko,
Nucl. Phys. B \textbf{312}, 509-550 (1989)
doi:10.1016/0550-3213(89)90570-1

\bibitem{Braun:1988qv}
V.~M.~Braun and I.~E.~Filyanov,
Z. Phys. C \textbf{44}, 157 (1989)
doi:10.1007/BF01548594

\bibitem{Chernyak:1990ag}
V.~L.~Chernyak and I.~R.~Zhitnitsky,
Nucl. Phys. B \textbf{345}, 137-172 (1990)
doi:10.1016/0550-3213(90)90612-H

\bibitem{Khodjamirian:2000mi}
A.~Khodjamirian,
Nucl. Phys. B \textbf{605}, 558-578 (2001)
doi:10.1016/S0550-3213(01)00194-8
[arXiv:hep-ph/0012271 [hep-ph]].

\bibitem{Khodjamirian:2003eq}
A.~Khodjamirian, T.~Mannel and B.~Melic,
Phys. Lett. B \textbf{571}, 75-84 (2003)
doi:10.1016/j.physletb.2003.08.012
[arXiv:hep-ph/0304179 [hep-ph]].

\bibitem{Khodjamirian:2017zdu}
A.~Khodjamirian and A.~A.~Petrov,
Phys. Lett. B \textbf{774}, 235-242 (2017)
doi:10.1016/j.physletb.2017.09.070
[arXiv:1706.07780 [hep-ph]].

\bibitem{Rusov:2017chr}
A.~V.~Rusov,
Eur. Phys. J. C \textbf{77}, no.7, 442 (2017)
doi:10.1140/epjc/s10052-017-5000-9
[arXiv:1705.01929 [hep-ph]].

\bibitem{Wang:2012kw}
Z.~G.~Wang,
Eur. Phys. J. A \textbf{49}, 131 (2013)
doi:10.1140/epja/i2013-13131-7
[arXiv:1203.6252 [hep-ph]].

\bibitem{Duplancic:2008ix}
G.~Duplancic, A.~Khodjamirian, T.~Mannel, B.~Melic and N.~Offen,
JHEP \textbf{04}, 014 (2008)
doi:10.1088/1126-6708/2008/04/014
[arXiv:0801.1796 [hep-ph]].

\bibitem{Khodjamirian:2020mlb}
A.~Khodjamirian, B.~Meli\'c, Y.~M.~Wang and Y.~B.~Wei,
JHEP \textbf{03}, 016 (2021)
doi:10.1007/JHEP03(2021)016
[arXiv:2011.11275 [hep-ph]].

\bibitem{ParticleDataGroup:2020ssz}
P.~A.~Zyla \textit{et al.} [Particle Data Group],
PTEP \textbf{2020}, no.8, 083C01 (2020)
doi:10.1093/ptep/ptaa104

\bibitem{ParticleDataGroup:2018ovx}
M.~Tanabashi \textit{et al.} [Particle Data Group],
Phys. Rev. D \textbf{98}, no.3, 030001 (2018)
doi:10.1103/PhysRevD.98.030001

\bibitem{Wang:2010it}
Z.~G.~Wang,
Eur. Phys. J. A \textbf{47}, 81 (2011)
doi:10.1140/epja/i2011-11081-8
[arXiv:1003.2838 [hep-ph]].

\bibitem{Roberts:2007ni}
W.~Roberts and M.~Pervin,
Int. J. Mod. Phys. A \textbf{23}, 2817-2860 (2008)
doi:10.1142/S0217751X08041219
[arXiv:0711.2492 [nucl-th]].

\bibitem{Hu:2017dzi}
X.~H.~Hu, Y.~L.~Shen, W.~Wang and Z.~X.~Zhao,
Chin. Phys. C \textbf{42}, no.12, 123102 (2018)
doi:10.1088/1674-1137/42/12/123102
[arXiv:1711.10289 [hep-ph]].


\bibitem{Wang:2010fq}
Z.~G.~Wang,
Eur. Phys. J. C \textbf{68}, 479-486 (2010)
doi:10.1140/epjc/s10052-010-1365-8
[arXiv:1001.1652 [hep-ph]].

\bibitem{Ball:1991bs}
P.~Ball, V.~M.~Braun and H.~G.~Dosch,
Phys. Rev. D \textbf{44}, 3567-3581 (1991)
doi:10.1103/PhysRevD.44.3567

\end{thebibliography}
\end{document}